\documentclass[11pt]{article}

\usepackage{fullpage}

\usepackage{graphics}
\usepackage{tikz}
\usetikzlibrary{arrows}
\usetikzlibrary{decorations.markings}
\usepackage{bbding}
\usepackage{subfig}
\usepackage{float}

\usepackage{latexsym}
\usepackage{amssymb}   

\usepackage{amsmath}

\usepackage{algpseudocode}
\usepackage{algorithm}

\usepackage{enumerate}
\usepackage{mathtools}
\usepackage{verbatim}
\usepackage{color}

\def\N{\mbox{$\Bbb N$}}
\def\K{\mbox{${\kappa}$}}
\def\R{\Bbb R}
\def\H{\mbox{$\cal H$}} 
\def\G{\mbox{$\cal G$}}
\def\F{\Bbb F}
\def\e{{{\epsilon}}}

\def\floor#1{\left\lfloor #1 \right\rfloor}
\def\finito{{\hspace*{\fill}  \mbox{$\blacksquare $}}}
\def\lemfinito{{\hspace*{\fill}  \mbox{$\blacktriangledown $}}}
\def\finito{{\hspace*{\fill}  \mbox{$\blacksquare $}}}

\newtheorem{theorem}{Theorem}[section]
\newtheorem{lemma}[theorem]{Lemma}
\newtheorem{definition}[theorem]{Definition}
\newtheorem{corollary}[theorem]{Corollary}
\newtheorem{conjecture}[theorem]{Conjecture}
\newtheorem{claim}[theorem]{Claim}
\newtheorem{question}[theorem]{Question}
\newtheorem{remark}[theorem]{Remark}

\begin{document}

\tikzset{->-/.style={decoration={
  markings,
  mark=at position .5 with {\arrow{>}}},postaction={decorate}}}

\tikzset{-<-/.style={decoration={
  markings,
  mark=at position .5 with {\arrow{<}}},postaction={decorate}}}

\author{
Louay Bazzi
\footnote{
Department of Electrical and Computer Engineering, American University of Beirut, 
Beirut, Lebanon. E-mail: louay.bazzi@aub.edu.lb.} 
\and 
Hani  Audah
\footnote{   
Department of Electrical and Computer Engineering, American University of Beirut, 
Beirut, Lebanon. E-mail: hta12@mail.aub.edu.lb} 
}
\title{Impact of redundant checks on the  LP decoding thresholds of LDPC codes\footnote{
Research supported by FEA URB grant, American University of Beirut.
}} 

\maketitle


\begin{abstract}
Feldman {\em et al.}  (2005) asked whether the performance of the Linear Programming (LP) decoder can be improved by adding   redundant parity checks to  tighten the LP relaxation.  We prove in this paper that for LDPC codes, even if we include  all redundant parity checks, asymptotically  there is no gain in the LP decoder threshold on the Binary Symmetric Channel  (BSC)  under certain conditions on the base Tanner graph. First,  we show that  if the Tanner graph has bounded check-degree and satisfies a condition which we call {\em asymptotic strength}, then including  high degree redundant parity checks in the LP   does not significantly improve the threshold of the LP decoder in the following  sense: for each constant $\delta>0$, there is a constant $k>0$ such that the threshold of the  LP decoder containing all redundant checks of degree at most $k$ improves by at most $\delta$ upon adding to the LP all  redundant checks of degree larger than $k$. We conclude that if the graph  satisfies an additional condition which we call {\em rigidity}, then including all   redundant checks does not improve the threshold of the base LP. We call the graph  {\em asymptotically strong} if  the LP decoder corrects a constant fraction of errors even  if the log-likelihood-ratios of the correct variables are arbitrarily small.  By building on a construction due Feldman {\em et al.}  (2007) and its recent improvement by Viderman (2013),  we show that asymptotic strength follows from sufficiently large  variable-to-check expansion. We also give a geometric interpretation of asymptotic strength in terms pseudocodewords. We call the graph {\em rigid} if 
the minimum weight of a sum of check nodes involving a cycle tends to infinity as  the block length  tends to infinity.  
Under the assumptions that the graph girth is logarithmic   and the minimum check degree is at least $3$, rigidity is equivalent to the {\em nondegeneracy} 
 property that adding at least logarithmically many checks does not give a constant weight check. We argue that nondegeneracy is a typical property of  random check-regular  Tanner graphs.
\end{abstract}

\newpage

\section{Introduction}\label{introS}

A  Low Density Parity Check (LDPC) code is a linear code whose parity check matrix  is sparse. 
LDPC codes were discovered by 
Gallager \cite{Gal62} in 1962  who 
used the  sparsity of the parity check  matrix to design   various  iterative decoding algorithms with good performance. 
 The parity check matrix of a LDPC is represented by 
a bipartite graph, called a Tanner graph \cite{Tan81}, between  a set of variable nodes and set of check nodes. 
The past two  decades saw  a growing number of research  results related  to LDPC codes and their iterative decoding algorithms 
(see \cite{RU08} for a comprehensive account). Graph properties such as good girth \cite{Gal62,Tan81} and expansion \cite{SS96} play a central role in 
designing good LDPC codes with efficient iterative decoding algorithms. 

  Linear Programming (LP) decoding of linear codes was introduced by Feldman {\em et al.}   
\cite{Fel03, FWK05} as a good-performance low-complexity relaxation of  Maximum Likelihood (ML) decoding.
In the past decade,  the good performance  of LP decoding of LDPC codes was established in a sequence of papers which
lead again to good girth and expansion as desirable properties of the underlying Tanner graph. 
The LP decoder corrects a constant fraction of errors  
if the graph has  sufficiently large expansion  \cite{FMS07,DDKW08, Vid13}. 
Moreover, the LP decoder of certain expander  codes  
achieves the capacity of a  wide class of  binary-input memoryless symmetric channels \cite{FS05}. 
Lower bounds on the LP decoding  thresholds  of LDPC codes where obtained in \cite{KV06,ADS12} 
under the assumption that  the graph has a logarithmic girth, and upper bounds were obtained in 
\cite{VK06}. 
The  LP decoding polytope was independently discovered by Koetter and Vontobel \cite{KV03} in the context of graph covers of Tanner graphs  
and iterative decoding algorithms. The link between LP decoding and 
iterative decoding algorithms,  in particular the min-sum algorithm, was further investigated 
in \cite{VK04,ADS12}.

Feldman {\em et al.}  \cite{Fel03,FWK05} asked whether the performance of the LP decoder can be improved by tightening the 
LP relaxation. Namely, they proposed two natural approaches  to tighten the LP: (1) adding {\em redundant parity checks} and 
(2)   {\em Lifting  techniques}.  Another tightening technique  based  on merging nodes was explored by  Burshtein and Goldenberg \cite{BG11}.

This paper is about the first approach. 
Including redundant parity checks does not affect the code but adds new constraints to the LP. 
The problem of appropriately selecting  redundant checks to be added to the LP without 
sacrificing its efficiency was  investigated in \cite{TS08,MWT09}.
Even though simulation results  suggest that redundant checks improve the LP decoder performance \cite{FWK05,TS08,MWT09}, 
 we argue in this paper that
   asymptotically 
   there is no gain in terms of the LP decoder threshold on the BSC 
 even if we add all redundant checks,    
assuming  that the base Tanner graph has bounded check-degree and satisfies two natural conditions which we call asymptotic  strength and rigidity.  
The required conditions are satisfied if in addition to sufficiently  good expansion and girth, the graph has a nondegeneracy property, which  
holds with high probability for   random check-regular graphs.

As for the lifting techniques,  a recent result  of  Ghazi and Lee \cite{GL14} shows  that extensions of the LP decoder based on  Sherali-Adams and Lasserre hierarchies  
do not significantly improve the error correction capabilities of  the LP decoder if the graph is a good expander.

The common theme between our result and the result of \cite{GL14}  is that if the 
base LP has ``certain desirable or typical properties'' then it is ``hard to make it asymptotically better''.  Related to this theme is the 
other extreme of {\em geometrically perfect codes}, which are by definition 
codes  for which the LP resulting from adding all redundant checks  is equivalent to ML decoding (see Section \ref{lpds}); 
such codes are asymptotically bad  by a recent result due to   Kashyap \cite{Kas08}.   

On the positive side,  
our negative results  suggest studying the  LP decoding limits  
in the framework of the dual code containing all redundant check nodes. 
This framework is appealing since it is independent of a particular Tanner graph representation of the  code.

The proof of our main result is   based on a careful  analysis of the dual LP. We use  the dual witness and hyperflow structures 
  \cite{FMS07,DDKW08} and the fact that the existence of such structures is necessary  for LP decoding success \cite{BGU14}. 
We also  use the the notion of acyclic hyperflows  and the LP excess technique  \cite{BGU14}.  
To establish the  relation between asymptotic strength and expansion, we build on the dual witness construction in   \cite{FMS07, Vid13}.
Our probabilistic analysis of  nondegeneracy  is based on the work of Calkin \cite{calkin}.

In the remainder of this   introductory section,  
we give background material on 
Tanner graphs, redundant checks  and LP decoding.  
Then, we formally state  our  results in Section \ref{sres} and  we give a detailed outline of the rest of the paper in Section \ref{outline}.

\subsection{Tanner graphs and redundant checks}\label{tanred}

A {\em Tanner graph}  $G=(V,C,E)$ is an undirected  bipartite graph between a set $V$ of {\em variable nodes} and 
a set $C$ of {\em check nodes}, where $E$ is the set of edges. 
If $i \in V$ is a variable node,  we will denote by $N(i)$ the check neighborhood of $i$, i.e., the set of check nodes adjacent to $i$.
Similarly, if $j \in C$ is check node,  $N(j)$ is  the set of variable  nodes adjacent to $j$.
Unless otherwise specified, we  assume throughout the paper   that $V = \{1,\ldots, n\}$, where $n\geq 1$ is the {\em block length}.
We assume also that the degree of each check node is at least one. 
The {\em linear code} $Q = Q_G$ associated with $G$ is the $\F_2$-linear code $Q\subset \F_2^n$ whose 
parity check matrix is the biadjacency matrix of $G$.  That is, 
$Q$ is the set of all binary strings $x\in \F_2^n$ such that $\sum_{i\in N(j)} x_i = 0$ for each $j \in C$.

Given a tanner graph $G=(V,C,E)$, the  Tanner {\em graph of all redundant checks} $\overline{G}$ associated with $G$ is defined 
as follows. 
A {\em redundant  check} of $G$ is a nonzero $\F_2$-linear combinations of  checks of $G$, thus 
the  redundant  checks  are in one-to-one correspondence with the nonzero  elements of the dual code $Q^\bot$.   The graph 
$\overline{G}$ is obtained from $G$ by adding  all redundant checks to $G$.
That is, $\overline{G} = (V, \overline{C}, \overline{E})$, where 
$\overline{C}= Q^\bot- \{0\}$ and  $i\in V$ is connected to $c\in Q^{\bot}$ iff  $c_{i}=1$.

 We are also interested in the following graded subgraphs of $\overline{G}$. 
Given $G = (V,C,E)$ and an integer $k$, let $\overline{G}^{k}$ be the  Tanner {\em graph 
 of redundant checks of degree at most $k$}. That is,  
$\overline{G}^{k} = (V,\overline{C}^k, \overline{E}^k)$  is the subgraph of $\overline{G}$ induced on $V$ and the set 
$\overline{C}^k$ of  nonzero checks of degree at most $k$, i.e., 
$\overline{C}^k = \{  c\neq 0 \in  Q^\bot: weight(c)\leq k\}$. 
Thus,   if $d$ is the maximum degree of a check node in $G$, 
we have the nested sequence of Tanner graphs 
$G \subset \overline{G}^{d} \subset \overline{G}^{d+1} \subset \ldots \subset \overline{G}^{n } = \overline{G}$, all defining the same code $Q$. 
Throughout this paper, we are in interested in base Tanner graphs  where  the {\em maximum check degree} $d$ is bounded.

\subsection{Linear programming decoder} \label{lpds}

Let $G = (V, C, E)$ be a   Tanner graph and $Q \subset \F_2^n$ the associated code.
Consider  transmitting a codeword of $Q$ over the 
the  $\epsilon$-{\em BSC $($Binary Symmetric Channel$)$}, which on input $x\in \F_2^n$ outputs  $y\in \F_2^n$ by 
flipping each bit of $x$ independently with probability $\epsilon$. 
The {\em ML $($Maximum Likelihood$)$} decoder is given by $\hat{x}_{\mathrm{ML}} = 
{\operatorname{argmax}}_{x\in Q}{~ p_{Y|X}(y|x)}$. 
Let $\gamma\in \R^n$ be the {\em LLR $($Log-Likelihood-Ratio$)$} vector of $y$: 
 $\gamma_i=\log{\big(\frac{p_{Y_{i}|X_{i}}(y_{i}|0)} {p_{Y_{i}|X_{i}}(y_{i}|1)}\big)} = (-1)^{y_i}\log{\frac{1-\epsilon}{\epsilon}}$ for $i=1,\ldots, n$.   
In terms of $\gamma$, the ML decoder is given by
\begin{equation}\label{mleq}
\hat{x}_{\mathrm{ML}} =\underset{x \in Q}{\operatorname{argmin}} \langle  x,\gamma\rangle,
\end{equation}
where $ \langle  x,\gamma  \rangle := \sum_ i x_ i \gamma_i$. 
For general linear codes,  the ML decoding problem is NP-hard \cite{BMVT78}.  Feldman {\em et al.}  \cite{Fel03, FWK05} 
introduced the 
approach  of {\em LP $($Linear Programming$)$} decoding, which is based on relaxing the  optimization problem on $Q$ into an LP. 
Due to  
the linearity of the objective function $ \langle  x,\gamma  \rangle $, optimizing over $Q$ is equivalent to optimizing over the 
convex polytope  $\operatorname{conv}(Q)  \subset \R^n$ spanned by the convex combinations of the codewords in $Q$:
\begin{equation}\label{mleq2}
\hat{x}_{\mathrm{ML}} = \underset{x \in \operatorname{conv}(Q)}{\operatorname{argmin}}\langle  x,\gamma \rangle. 
\end{equation}
 The idea of Feldman is to relax $\operatorname{conv}(Q)$ into a larger  lower-complexity   polytope.   
For each check node $j\in C$, define the local code $Q_j$ consisting of all vectors $x \in \{0,1\}^{n}$ satisfying check $j$, thus 
 $Q =  \bigcap_{j\in C} Q_{j}$.  Let 
\begin{equation}\label{fundpo}
  P(G) := \underset{j \in C} \bigcap \operatorname{conv}(Q_{j}) \supset \operatorname{conv}(\underset{j \in C} \bigcap Q_{j}) = \operatorname{conv}(Q). 
\end{equation}  
The polytope $P(G)$ depends on the Tanner graph representation of the code and it is called the 
{\em fundamental polytope} of $G$.  The {\em LP decoder} is the relaxation of  the ML decoder given by 
\begin{equation}\label{le:relaxed_LP}
\hat{x}_{LP} = \underset{x \in P(G)}{\operatorname{argmin}}\langle  x,\gamma \rangle.
\end{equation}
The relaxed LP can be efficiently solved due to the low complexity of $P(G)$. 
More generally, (\ref{mleq}) and (\ref{le:relaxed_LP}) define 
the ML and LP decoder   for an arbitrary LLR vector $\gamma\in \R^n$. 
If $\gamma$ is as above associated with  a binary vector $y$, 
we ignore without loss of generality 
the constant $\log{\frac{1-\epsilon}{\epsilon}}$ and we  
normalize $\gamma$ so that  $\gamma =(-1)^y$.

It is appropriate to mention at this stage geometrically perfect codes. A linear code $Q\subset \F_2^n$ is called 
{\em geometrically perfect} \cite{BG86,Kas08} if the LP relaxation 
corresponding to the full dual code is exact, i.e., $P(\overline{G}) = conv(Q)$, where $G$ is any Tanner graph of $Q$. 
Examples of such codes are  tree codes and cycle codes. 
Geometrically perfect codes  are classified in \cite{BG86} based on  Seymour’s matroid decomposition theory \cite{Sey80},  
but they are unfortunately asymptotically bad in the sense that 
their minimum distance does not grow linearly with the  block length \cite{Kas08}.

We are interested in   LP thresholds over the BSC as the block length $n$ tends to infinity. 
That is, we have an {\em infinite family of Tanner graphs} $\G = \{G_n \}_n$, where $G_n = (V_n,C_n,E_n)$ is  a Tanner  
graph on $n$ variable nodes, i.e., $V_n = \{ 1,\ldots, n\}$. 
Define the {\em LP-threshold}  $\xi_{LP}(\G)$ of $\G$ to be 
the supremum of $\e\geq 0$ such that the error probability of the  LP decoder of $G_n$ over the $\e$-BSC goes to zero as $n$ tends to infinity, i.e.,  
\[
   \xi_{LP}(\G) = \sup\{\e\geq 0   ~:~  \operatorname{Pr}_{\e\mbox{-}BSC}[\mbox{ LP decoder of $G_n$  fails}] = o(1) \}.  
\]
As in previous work \cite{FWK05}, 
we assume without loss of generality  that the all-zeros codeword was 
transmitted and that the LP decoder fails if zero is not the unique optimal solution of the LP. 

Finally, given an infinite family of Tanner graphs $\G = \{G_n \}_n$, we are interested 
in the resulting family   $\overline{\G} := \{ \overline{G_n}\}_n$ of Tanner graphs obtained by adding all redundant checks.
Moreover, if $k: \N^+ \rightarrow \R$,   
 we are also interested in the  family   $\overline{\G}^k := \{ \overline{G_n}^{k(n)}\}_n$ of Tanner graphs obtained by  adding all redundant checks 
of degree at most $k$.

\subsection{Summary of results}\label{sres}

Let $\G = \{G_n\}_n$ be an infinite family of Tanner graphs of bounded check degree. We show that 
if $\G$ satisfies a condition which we call {\em asymptotic strength}, then including high degree redundant checks in the LP 
 does not improve the threshold in the sense that 
for each constant $\delta>0$, there is a constant $k>0$ such that $\xi_{LP}(\overline{\G}^k) \geq \xi_{LP}(\overline{\G}) - \delta$. 
We conclude that if $\G$  satisfies an additional condition which we call {\em rigidity}, 
then including all   redundant checks does not improve the threshold of the base LP 
in the sense that  $\xi_{LP}(\G) = \xi_{LP}(\overline{\G})$. 
We call the graph {\em asymptotically strong} if  the LP decoder corrects a constant fraction of errors even 
if the LLR values of the correct variables are arbitrarily small.   
 We show that the asymptotic strength condition follows from expansion. 
We call the graph {\em rigid} if  
the minimum weight of a sum of check nodes involving a cycle tends to infinity as  the block length  tends to infinity.  
We note that under the assumptions that the girth of $G_n$ is $\Theta(\log{n})$ and the minimum check degree is at least $3$, rigidity  is 
equivalent to the property that adding $\Omega(\log{n})$ checks does not give $O(1)$ weight checks, which we argue is 
a typical property of  random check-regular  Tanner graphs.

\begin{definition}[Asymptotically strong Tanner graphs]\label{asstrong} Let  
$\G = \{G_n\}_n$ be an  infinite  family of Tanner graphs. 
We call $\G$ {\em asymptotically strong} if for each  (small) constant $\beta>0$, there exists a constant 
$\alpha>0$ such that for each $n$ and each error vector  $y\in \{0,1\}^n$ of weight at most $\alpha n$, 
 the LP decoder of $G_n$ succeeds on the asymmetric LLR vector $\gamma({y,\beta})\in  \R^n$ given by 
\[                        
\gamma_i({y,\beta}) = \left\{\begin{array}{ll} -1 &\mbox{ if } y_i=1  \\ \beta & \mbox{ if } y_i=0,\end{array}\right.
\]
for  $i=1,\ldots,n$.  
\end{definition}
Although  asymmetry in the LLR vector might seem  unnatural at this point, 
we start with this definition of asymptotic strength because 
it gives flexibility in the analysis. We give later an equivalent definition  in terms 
of pseudocodewords (Theorem \ref{asseq}). 

\begin{theorem}[High degree  redundant checks do not improve LP threshold]\label{crr1}
~~Let  $\G =$\\$\{G_n\}_n$ be an  infinite   family of  Tanner graphs such that each check node has degree at most $d$, where $d$ is a constant.  
Assume that  $\G$ is asymptotically strong. 
 Then:
\begin{itemize}
\item[a)] For any small constant  $\delta>0$, there exists a sufficiently large constant $k\geq d$ 
$($dependent on $\delta$ and 
independent of $n$$)$
  such that 
       $\xi_{LP}(\overline{\G}^k)\geq  \xi_{LP}(\overline{\G}) - \delta$.   
\item[b)] If  $k(n)$  is a real valued function of $n$ such that $k(n) = \omega(1)$ $($i.e., $k(n)$ tends to infinity as $n$ tends to infinity$)$, 
then $\xi_{LP}(\overline{\G}^{k})=  \xi_{LP}(\overline{\G})$.
\end{itemize}
\end{theorem}
The proof of Theorem \ref{crr1} uses the LP excess lemma \cite{BGU14} and the notion of primitive hyperflows which we define at the end of this section.

Feldman {\em et al.}  \cite{FMS07} argued that expansion implies that 
the LP decoder corrects a positive fraction of errors.  
The link between the expansion of a Tanner graph and the error correction capabilities of the underlying code 
 was discovered by  Sipser and Spielman \cite{SS96} in the context of iterative decoding algorithms.
Recently, Viderman \cite{Vid13} simplified the argument of \cite{FMS07} and improved its dependency on the 
expansion parameter.  
By building on  the construction in \cite{FMS07,Vid13}, we show that graphs with good expansion are 
asymptotically strong.

A Tanner graph $G=(V,C,E)$ is called an {\em $(\varepsilon n, \K)$-expander} if for each subset $S \subset V$ of variable nodes of size at most 
$\varepsilon n$, 
we have $|N(S)| \geq \K |S|$, where $N(S)$ is the set of (check) nodes adjacent to $S$. 

\begin{theorem} [Expansion implies  asymptotic strength]\label{expstr}
Let $d_v>0,\varepsilon>0$ and  $\delta > \frac{2}{3}$ be constants such that $d_v$ is an integer and 
  $\delta d_v$ is an integer. 
Let $\G = \{G_n\}_n$ be an  infinite   family of   Tanner graphs with regular variable degree $d_v$  and bounded check degree. 
If $G_n$ is an  $(\varepsilon n, \delta d_v)$-expander for each $n$,   
then $\G$ is asymptotically strong. 
\end{theorem}
It is known that  redundant check nodes  obtained  by acyclic sums of check nodes do not tighten the polytope 
\cite{Fel03,KVLarge,BG11}, which motivates the following definitions.

\begin{definition}[Cylic sums of checks]\label{essred}
Let $G = (V,C,E)$ be a Tanner graph .
We call a subset of check nodes   $S\subset C$ {\bf cyclic} if the graph induced by $G$ on $S$ contains a cycle.

Define $\Delta(G)$ to be the {\bf minimum  weight of the sum of a cyclic subset of check nodes of $G$}.
More formally, let $Q \subset \F_2^n$ be the code associated with $G$. 
For each check $j \in C$, 
let $z_j\in Q^\bot$  be   the vector in the dual code associated with $j$.  Then 
\[
  \Delta(G) := \min\{ weight(\sum_{j \in S} z_j) ~:~ S \subset C~\mbox{cyclic} \}. 
\] 
\end{definition}
\begin{definition}[Rigid Tanner graphs]\label{essred2}
We call an infinite family $\G=\{G_n\}_n$ of Tanner graphs {\bf rigid } if  the minimum weight  of a sum of check nodes involving a cycle tends to infinity as  the block length  tends to infinity.  More formally, $\G$ is rigid if $\Delta(G_n)=\omega(1)$.
\end{definition}
\begin{remark} {\em 
If $\G$ is rigid, then the check nodes of $G_n$ are linearly independent for 
sufficiently large $n$ (since any subset of check nodes whose sum is zero must be cyclic).
}
\end{remark}

Accordingly, we obtain the following corollary to Theorem \ref{crr1}.  
\begin{corollary}[Redundant checks do not improve LP threshold] \label{crr2.2}
Let $\G=\{G_n\}_n$ be an  infinite   family of Tanner graphs of bounded check degree. 
If $\G$ is asymptotically strong and rigid, then $\xi_{LP}(\overline{\G}) = \xi_{LP}(\G)$. 
\end{corollary}
It is not hard to see that  $\omega(1)$-girth is a necessary condition for rigidity.  
Unfortunately, random graphs have $O(1)$-girth, thus they are not necessarily rigid. 
In general, $\Theta(\log{n})$-girth is a desirable property of a Tanner graph in the context of LP decoding \cite{FWK05}
and iterative decoding \cite{Gal62,Tan81}. Random graphs with good girth are typically constructed by breaking the cycles of a random graph. 
We note that for graphs with $\Theta(\log{n})$-girth and minimum check degree at least $3$, rigidity is equivalent to 
 a simpler nondegeneracy condition which we define below.

\begin{definition} [Nondegeneracy]\label{mincycw} 
Call an $m \times n$ matrix  $M \in \F_2^{m\times n}$ 
$(s,k)$-nondegenerate  
if the sum of any subset of at least $s$ rows of $M$ has weight larger than $k$.
We call a Tanner graph $G$  $(s,k)$-nondegenerate if its $m\times n$ biadjacency matrix is $(s,k)$-nondegenerate, where 
$m$ is the number of check nodes and $n$ is the number of variable nodes. 
\end{definition}
For instance, full row rank corresponds to  $(1,0)$-nondegeneracy.

\begin{lemma}[Rigidity versus girth and nondegeneracy]\label{lcrr3}
Let $\G=\{G_n\}_n$ be an  infinite   family of Tanner graphs of bounded check degree.     
If $\G$ is rigid, then $girth(G_n) = \omega(1)$. 
On the other hand,  if  $girth(G_n) = \Theta(\log{n})$  and the minimum check degree of $\G$ is at least $3$ (i.e., for all $n$, each check node of $G_n$ has degree at least $3$), then the following are equivalent: 
\begin{itemize}
\item[i)] {\em (Rigidity)} $\G$ is rigid 
\item[ii)] {\em (Nondegeneracy)} For each constant $c >0$, $G_n$ is $(c \log{n}, \omega(1))$-nondegenerate. \\
That is,  for each constant $c>0$, the minimum weight of a sum of at least $c \log{n}$ checks nodes tends to infinity as $n$ increases.
\end{itemize}
\end{lemma}  
We argue   that  nondegeneracy is a typical property of random check-regular Tanner graphs. 
Namely, we  show that random check-regular graphs are $(c\log{n}, \omega(1))$-nondegenerate with high probability if 
$m\leq \beta_d n$, where $d$ 
the check degree and and $\beta_d$ is Calkin's threshold as given in  Definition \ref{calkinthr} 
$($$\beta_d$ is a threshold close to $1$, e.g.,  $\beta_3 \sim 0.8895$,  $\beta_4 \sim 0.967$ and $\beta_5 \sim 0.989$$)$. 
\begin{lemma} [Random check-regular graphs are nondegenerate]\label{randlem}
Let $d,m$ and $n$ be integers such that $d\geq 3$ and  $1\leq m < \beta_d n$.
Consider a random $m\times n$ matrix $M\in \F_2^{m\times n}$ constructed  by independently 
choosing each of the  $m$ rows of $M$ uniformly from  the set of vectors in $\F_2^n$ of weight $d$. 
Then for any constant $c>0$ and any function $k(n)$ of $n$ such that  $k(n) = o(\log{\log{n}})$, 
$M$ is $(c\log{n}, k(n))$-nondegenerate with high probability.
\end{lemma}
We establish the claim by adapting an argument used by Calkin \cite{calkin} to show that if $m < \beta_d n$, then $M$ has full row rank with high probability. 
The ensemble of  random check-regular graphs   is attractive from a  probabilistic analysis standpoint, but  it 
 typically gives  irregular graphs with constant girth.
 We believe that good girth and variable-regularity do not increase the odds  of degeneracy;  
we conjecture that the statement of Lemma \ref{randlem} extends to the ensemble of regular  $\Theta(\log{n})$-girth  Tanner graphs   
(see Section \ref{disop}).

We also prove the following general results about LP decoding which might be of independent interest:  
\begin{itemize}
\item {(\bf Primitive hyperflows})  We give a simple necessary and sufficient condition for the success of the LP decoding  when all  redundant checks are included in the LP. 
The condition is in terms of the existence of a hyperflow (see Definition \ref{dualdefs}) which 
is primitive in the sense that all the variables in error have zero outflow and all the correct variables have zero inflow (Theorem \ref{primlem}). 
This characterization is essential to the proof of Theorem \ref{crr1}.  
\item 
{(\bf Pseudocodewords interpretation of asymptotic strength})
We note that the notion of asymptotic strength 
has the following geometric interpretation in terms of 
pseudocodewords:
 $\G=\{G_n\}_n$ is asymptotically strong iff for each nonzero pseudocodeword $x \in P(G_n)$, to attain a  positive fraction of $\sum_i x_i$, we need a least  linear number of 
coordinates of $x$. That is,  for each  $\theta>0$, there exists $\alpha>0$ such that for each $n$ and each nonzero pseudocodeword $x \in P(G_n)$,
the sum of the largest $\lfloor \alpha n \rfloor $ coordinates  of $x$ is less than $\theta\sum_i x_i$ 
(Theorem \ref{asseq}). 
\item  {(\bf Asymptotic strength and LP decoding with help}) 
Assume that we are allowed to 
to flip at most a certain number of bits of the corrupted codeword  to help 
the LP decoder on the BSC.
We argue that if the Tanner graph is asymptotically strong, allowing 
a sublinear number of help bits does not improve the LP threshold (Theorem \ref{tbcp2}).
This result,  although a negative statement, has potential constructive applications as it weakens the dual witness 
requirement for LP decoding success. 
\item {(\bf LP deficiency lemma})
We give a converse of the LP excess lemma \cite{BGU14}. Namely,  we show how to trade LP-deficiency 
for crossover probability (Lemma \ref{defilemma}) and  we use the LP deficiency lemma  
 to establish 
the above result on LP decoding with help. 
\end{itemize}

\subsection{Outline}\label{outline}

In Section \ref{le:LPandgraphs}, we give background material on graph structures 
whose  existence is necessary and sufficient for LP decoding success:
dual witness, hyperflows and acyclic hyperflows. 
To warm up, we highlight in Section \ref{hdcs} a simple classical argument, which shows that 
high density codes have zero thresholds on the BSC.   
The key starting point of our proof is 
the above-mentioned  special type of hyperflows called   primitive hyperflows.
We define primitive hyperflows  in Section \ref{primhpf} and we argue that their existence is sufficient for LP decoding  success  when all  redundant checks are included in the LP. 
In Section \ref{impred}, we show that for asymptotically strong codes with bounded-check degree, high degree checks  do not  improve the threshold (Theorem \ref{crr1}). 
Then we conclude that 
adding all redundant checks does not improve the threshold  if the graph is additionally  rigid (Corollary \ref{crr2.2}). 
In Section \ref{expanstr}, we study the relation between expansion and asymptotic strength (Theorem \ref{expstr}). 
In Section \ref{randSec}, we study the rigidity and the related  nondegeneracy properties (Lemmas \ref{lcrr3} and \ref{randlem}).
In Section \ref{pseudoint}, we give  the above-mentioned pseudocodewords  interpretation of asymptotic strength. 
In Section \ref{dechelp}, we give an application of asymptotically strong codes in the context of the above-mentioned problem of 
 LP decoding with help bits.  
Finally, we conclude in Section \ref{disop}  with a discussion  of the asymptotic strength condition, the  rigidity condition  and the limits of LP decoding 
on the BSC.

\section{LP decoding success, dual witness and  hyperflow}\label{le:LPandgraphs}

In this section we summarize various dual  characterizations of 
 LP decoding success that will be used in this paper.    
The notion of dual witness  was introduced in \cite{FMS07} as a sufficient condition for LP decoding success. 
The necessity of the existence of a dual witness  for LP decoding success was established in  \cite{BGU14}. 
A special type of  dual witnesses called hyperflows 
was introduced in \cite{FMS07,DDKW08}. The equivalence between 
the existence of a hyperflow and  the existence 
of a dual witness was established in \cite{DDKW08}. 
The notion of a hyperflow  was further simplified in \cite{BGU14} who argued that the 
the existence of an acyclic hyperflow is equivalent to the existence of a hyperflow.

\begin{definition}[\cite{FMS07,DDKW08}]\label{dualdefs} 
{\em \bf (Dual witness, Hyperflow, and WDG)} 
 Consider a Tanner graph $G=(V,C,E)$ and an LLR vector $\gamma\in \R^V$. 
A {\bf dual witness}  for $\gamma$ in $G$ is a function $w: E \to \mathbb{R}$
 satisfying the inequalities in (a) and (b) below.  
\begin{itemize}
\item[a)]  {\bf Variable nodes inequalities:} $F_i(w)  <  \gamma_i$, for each variable $i\in V$,  where
$F(w)\in \R^V$ is given by 
 $$F_i(w) := \sum_{j \in N(i)} w(i,j).$$
We call $F_i(w)$ the flow  at variable node $i$ associated with $w$. 
\item[b)] {\bf Check nodes inequalities:} for each check $j\in C$  and all distinct variables $i\neq i' \in N(j)$, $w(i,j) + w(i',j) \geq 0$. 
\end{itemize}
A dual witness $w: E \to \mathbb{R}$ is called a {\bf hyperflow}  
if, instead of (b), it satisfies 
the following stronger check nodes inequalities. 
\begin{itemize}
\item[c)] {\bf Hyperflow check nodes inequalities:} 
 for each check $j\in C$, there exists $P_j\geq 0$ and a variable $i\in N(j)$ such that 
 $w(i,j) = -P_{j}$ and  $w(i',j) = P_{j}$, for all $i'\neq i \in N(j)$.
\end{itemize}
A dual witness or a hyperflow  $w$ can viewed as a {\bf weighted directed graph (WDG)}  $D$ on the vertices $V \cup C$, where 
an arrow is directed from $i$ to $j$ if $w(i,j) > 0$, an arrow is directed from $j$ to $i$ if $w(i,j) < 0$ and $i$ and $j$ are not connected by an arrow if $w(i,j) = 0$. 
The weight of each  directed edge connecting $i \in V$ and $j \in C$ is  $|w(i,j)|$. Thus, in terms of $D$, the variable nodes inequalities in (a) can be rephrased 
as follows.  
 \begin{itemize}
\item[d)]  {\bf WDG variable nodes inequalities:} 
  $F_i^{out}(w)  <  F_i^{in}(w)+ \gamma_i$
, for each variable $i\in V$, 
  where $F^{out}(w),  F^{in}(w) \in \R^V$ are defined as follows. 
\begin{itemize}
\item[$\bullet$]
$F_i^{out}(w)
 :=    \sum_{j\in Out_D(i)} |w(i,j)|$ where 
 $Out_D(i)$ is the  set of check nodes incident to  edges outgoing from $i$. 
\item[$\bullet$] 
$F_i^{in}(w) :=   \sum_{j\in In_D(i)} |w(j,i)|$ where 
and $In_D(i)$ is the  set of check nodes incident to edges ingoing to $i$. 
\end{itemize}  
We call $F_i^{out}(w)$ the   outflow from variable node $i$ associated with $w$ and 
$F_i^{in}(w)$ the   inflow to variable node $i$ associated with $w$.   
\end{itemize}
\end{definition}

We summarize in the following  theorem various equivalent  characterizations of LP decoding success.

\begin{theorem}[\cite{FMS07,DDKW08, BGU14}]\label{equivChars}
{\em\bf (Equivalent characterizations of LP decoding success )}
 Let $G=(V,C,E)$ be a Tanner graph  and $\gamma\in  \R^V$ an  LLR vector. 
   Then the following are equivalent:
\begin{enumerate}
\item[i)] The LP decoder of $G$ succeeds on $\gamma$ $($i.e., it returns zero as the unique solution under the assumption that the all-zeros codeword was transmitted$)$.
\item[ii)] There is a dual witness for $\gamma$ in $G$. 
\item[iii)] There is a hyperflow for $\gamma$ in $G$. 
\item[iv)] There is a hyperflow for $\gamma$ in $G$  whose  WDG is acyclic. 
\end{enumerate}
\end{theorem}
\begin{remark} {\em 
The fact that (ii) implies (i) follows from  \cite{FMS07}, 
the fact that  (i) implies (ii) follows from  Theorem 3.2 and  Remark 3.3 in \cite{BGU14}, 
 the equivalence between (ii) and (iii) follows from  Proposition $1$ in \cite{DDKW08} and the 
 the equivalence between (ii) and (iv) follows from  Theorem 3.7 in \cite{BGU14}. 
Note that  the statement of Theorem 3.7 in \cite{BGU14} assumes that  $\gamma$ is an
 LLR vector of a binary error pattern (i.e., $\gamma\in \{ -1,1\}^V$), but its proof 
holds for an arbitrary  LLR vector $\gamma\in  \R^V$. }
\end{remark}

\section{High density codes}\label{hdcs}

In this section, we highlight   a simple classical argument which shows that  high density codes have zero thresholds on the BSC.  
 A statement similar to Lemma \ref{highd} below   appears in Corollary 7  of \cite{VK06} in the context of regular Tanner graphs (with a different but also simple proof). 
Although not used   in  the proofs of the results in this paper,  we include this lemma  since from a broad perspective  
it is related to  the statement of Theorem \ref{crr1}, which says that high degree redundant checks are not 
helpful if the code is asymptotically strong. Unfortunately, the  simple proof of Lemma \ref{highd} does not extend to the setup of high degree redundant checks.

\begin{lemma}[High density codes] \label{highd}
Let $G=(V,C,E)$ be a Tanner graph such that the minimum degree of a check node is  $d_{min}$.
Then the LP decoder of $G$ fails if the number of errors introduced by the BSC 
is at least $n/d_{min}$. 
Thus, if  $\G$ is an infinite family of  Tanner graphs such that the minimum degree of a check node in $G_n$ is $\omega(1)$, then the threshold  $\xi_{LP}(\G)=0$.
\end{lemma}  
{\bf Proof:}
Assume that the all-zeros codeword was transmitted and let $y\in \{0,1\}^n$ 
be the received vector. 
 If the LP decoder of $G$ correctly decodes $y$, then 
by Theorem \ref{equivChars},
$(-1)^y$ has a hyperflow $w: E \rightarrow \R$. 
Consider the WDG $D$ corresponding to $w$ and 
let $U = \{i : y_i =1\}$ be the set of variables in error. 
If $S \subset V$, let 
$F^{in}(w;S) := \sum_{i \in S} F_i^{in}(w)$ be total inflow to $S$ and $F^{out}(w;S) := \sum_{i \in S} F_i^{out}(w)$ be total outflow from $S$. 
Summing the  variable nodes inequalities   $F_i^{out}(w)  <  F_i^{in}(w)+ \gamma_i$
over all $i\in V$, we get $F^{out}(w;V)<F^{in}(w;V) + |U^c|-|U|$, i.e., 
\begin{equation}\label{esfdf1}
F^{out}(w;V)<  F^{in}(w;V) + n-2|U|. 
\end{equation}
Summing the  variable nodes inequalities over all $i\in U$, we get 
$F^{out}(w;U)<F^{in}(w;U) -|U|$. Since 
$F^{out}(w;U)\geq 0$ and $F^{in}(w;U)\leq F^{in}(w;V)$, we obtain 
\begin{equation}\label{esfdf2}
|U|< F^{in}(w;V).
 \end{equation}
Finally, the hyperflow check nodes inequalities ((c) in Definition \ref{dualdefs}) imply that 
\begin{equation}\label{esfdf3}
(d_{min}-1)F^{in}(w;V) \leq F^{out}(w;V).
 \end{equation}
Solving for $|U|$ in (\ref{esfdf1}), (\ref{esfdf2}) and (\ref{esfdf3}), we obtain $|U|< n/d_{min}$. \finito

\section{Redundant checks and primitive hyperflows}\label{primhpf}

  We give in this section a  simple necessary and sufficient condition for the  success of LP decoding  when all redundant checks are included in the LP. 
The condition is in terms of the existence of a primitive hyperflow which we define as a hyperflow 
such that all the variables in error have zero outflow and all the correct variables have zero inflow. 
Primitive hyperflows are central to  the proof of Theorem \ref{crr1}.

\begin{definition}[Primitive hyperflow]
 Let $H=(V,C,E)$ be a Tanner graph,  $\gamma\in \R^V$ an LLR vector and  $w: E \to \mathbb{R}$ 
a hyperflow for $\gamma$ in $H$. 
Consider the WDG $D$ of $w$. 
We call $w$  {\bf a  primitive hyperflow} if for  each variable nodes $i\in V$, we have:  
\begin{itemize}
\item[a)]
If $\gamma_i\leq 0$, then $i$ has no outgoing edges in $D$, i.e., 
$F_{i}^{out}(w) = 0$. 
\item[b)] 
If $\gamma_i> 0$, then $i$ has no ingoing edges in $D$, i.e., $F_i^{in}(w) = 0$. 
\end{itemize}
Note that the WDG of a primitive hyperflow is necessarily  acyclic.  
\end{definition}

\begin{lemma}[Redundant checks and primitive hyperflows] \label{primlem}
 Let $G=(V,C,E)$ be a Tanner graph and consider the associated Tanner graph $\overline{G}=(V,\overline{C},\overline{E})$ of all redundant check nodes. 
Let   $\gamma\in \R^V$ be an  LLR vector. 
If the  LP decoder of $\overline{G}$    succeeds on $\gamma$, then  there is a primitive hyperflow for $\gamma$ in $\overline{G}$  
\end{lemma}  
{\bf Proof:} Assume that  the  LP decoder of $\overline{G}$   succeeds on $\gamma$. By Theorem \ref{equivChars},   there exists  a hyperflow $w:\overline{E}\rightarrow \R$ 
 for $\gamma$ in $\overline{G}$  
whose WDG $D$ is {\em acyclic}.  We will make $D$ primitive by exploiting the  key property of 
$\overline{G}$ that its check nodes are in one-to-one correspondence with the nonzero vectors in the dual $Q^\bot$
of the code $Q$ of $\overline{G}$. Hence,      
the $\F_2$-sum of any two distinct check nodes in  $\overline{G}$ is  again a check node in $\overline{G}$. 
 We will iteratively modify  $D$ until it  becomes   primitive by {\em repeated XORing} of check nodes. 
The basic operation is the   Switch operation in Algorithm \ref{switching}, which given 
a variable node  $i\in V$  and distinct check nodes  $j,j'\in C$ such 
that $(j,i)$ and  $(i,j')$ are edges in $D$,  modifies $D$ by replacing either $j$ or $j'$ with the XOR $j''$ of $j$ and $j'$. 
A key property of the Switch operation  is that it does not  increase  the indegree or the outdegree of $i$ and it decreases 
at least one of them.  
The Switch operation uses the fact that $D$ is acyclic.
\begin{algorithm}
\caption{Basic Switch operation} \label{switching}
{\bf Switch} $D$ along path $j\rightarrow i \rightarrow j'$ \\
Input: variable node $i\in V$  and check nodes $j,j'\in C$ such that $(j,i)$ and  $(i,j')$ are edges in $D$ 
\begin{algorithmic}[1]
\State  Let $P = \min\{|w(j,i)|,|w(i,j')|\}$
\State  Decrease  by $P$ the absolute weights of all the directed edges connected to $j$ or $j'$
\State  Let $i'$ be the (unique) variable  node such that $(j',i')$  is an edge in $D$ 
\State\label{L5}Let $j''$ be the XOR of $j$ and $j'$ 
\State\label{L6}Increase by $P$ the absolute weights the edges $(j'',i')$ and $(i'',j'')$, $\forall i''\neq i'\in N(j'')$
\State  Remove all zero weight edges.  
\end{algorithmic}
Figures \ref{switchingIl1}  and \ref{switchingIl2}  illustrate the Switch operation. 
\end{algorithm}

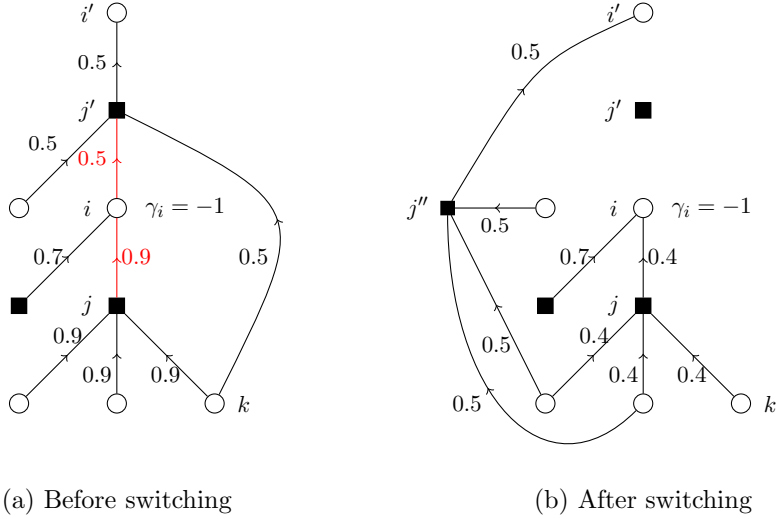
\begin{figure}[!h]
\centering
{
\subfloat{
\begin{tikzpicture}[scale=1.3] 
  \node[fill=none, scale=0.9, color=black]  () at (0,-1) {(a) Before switching}; 
  \node[draw, circle, scale=0.7] (a) at (-1,0) {}; 
  \node[draw, circle, scale=0.7] (b) at (0,0) {}; 
  \node[draw, circle, scale=0.7] (c) at (1,0) {}; 
  \node[draw, circle, scale=0.7] (f) at (0,2) {}; 
  \node[draw, circle, scale=0.7] (g) at (-1,2) {};
  \node[draw, circle, scale=0.7] (i) at (0,4) {}; 
  \node[draw, rectangle,fill=black, scale=0.8] (d) at (0,1) {}; 
  \node[draw, rectangle,fill=black, scale=0.8] (e) at (-1,1) {}; 
  \node[draw, rectangle,fill=black, scale=0.8] (h) at (0,3) {};

  \node[fill=none,color=black,scale=0.8] at (-0.3,1) {$j$};
  \node[fill=none,color=black,scale=.8] at (-0.3,2) {$i$};
  \node[fill=none,color=black,scale=.8] at (-0.3,3) {$j'$};
  \node[fill=none,color=black,scale=.8] at (-0.3,4) {$i'$};
  \node[fill=none,color=black,scale=.8]  at (0.7,2) {$\gamma_i=-1$};
  \node[fill=none,color=black,scale=.8] at (1.3,0) {$k$};

  \draw[->-,color=black] (a) --  (d);
  \draw[->-,color=black] (b) --  (d);
  \draw[->-,color=black] (c) --  (d);
  \draw[->-,color=red] (d) -- (f);
  \draw[->-,color=black] (e) -- (f);
  \draw[->-,color=red] (f) -- node[auto]{$\scriptstyle 0.5$}(h);
  \draw[->-,color=black] (g) -- node[auto]{$\scriptstyle 0.5$}(h);
  \draw[->-,color=black] (c).. controls (2,2) ..     (h);
  \draw[->-,color=black] (h) -- node[auto]{$\scriptstyle 0.5$} (i);
  \node[fill=none,color=black,scale=0.8] at (1.4,1.5) {$0.5$};
  \node[fill=none,color=black,scale=0.8] at (-0.2,0.3) {$0.9$};
  \node[fill=none,color=black,scale=0.8] at (-0.5,0.7) {$0.9$};
  \node[fill=none,color=black,scale=0.8] at (0.5,0.3) {$0.9$};
  \node[fill=none,color=red,scale=0.8] at (0.2,1.5) {$0.9$};
  \node[fill=none,color=black,scale=0.8] at (-0.7,1.5) {$0.7$};

\end{tikzpicture}}}
\qquad
{
\subfloat{
\begin{tikzpicture}[scale=1.3]
 \node[fill=none, scale=0.9, color=black]  () at (0,-1) {(b) After switching};
  \node[draw, circle, scale=0.7] (a) at (-1,0) {}; 
  \node[draw, circle, scale=0.7] (b) at (0,0) {}; 
  \node[draw, circle, scale=0.7] (c) at (1,0) {}; 
  \node[draw, circle, scale=0.7] (f) at (0,2) {}; 
  \node[draw, circle, scale=0.7] (g) at (-1,2) {};
  \node[draw, circle, scale=0.7] (i) at (0,4) {}; 
  \node[draw, rectangle,fill=black, scale=0.8] (d) at (0,1) {}; 
  \node[draw, rectangle,fill=black, scale=0.8] (e) at (-1,1) {}; 
  \node[draw, rectangle,fill=black, scale=0.8] (h) at (0,3) {};
  \node[draw, rectangle, fill=black,scale=0.7] (new) at (-2,2) {};

  \node[fill=none,color=black,scale=0.8] at (-0.3,1) {$j$};
  \node[fill=none,color=black,scale=.8] at (-0.3,2) {$i$};
  \node[fill=none,color=black,scale=.8] at (-0.3,3) {$j'$};
  \node[fill=none,color=black,scale=.8] at (-0.3,4) {$i'$};
  \node[fill=none,color=black,scale=.8]  at (0.7,2) {$\gamma_i=-1$};
  \node[fill=none,color=black,scale=.8] at (-2.3,2) {$j''$};
  \node[fill=none,color=black,scale=.8] at (1.3,0) {$k$};

  \draw[->-,color=black] (a) --  (d);
  \draw[->-,color=black] (b) -- (d);
  \draw[->-,color=black] (c) -- (d);
  \draw[->-,color=black] (d) -- (f);
  \draw[->-,color=black] (e) -- (f);
  \draw[->-,color=black] (a) --  (new);
  \draw[->-,color=black] (b) .. controls (-1,-1) and (-2:-2) .. (new);
  \node[fill=none,color=black,scale=0.8] at (-1.8,0) {$0.5$};
  \draw[->-,color=black] (g) --   node[auto]{$\scriptstyle 0.5$}(new);
  \node[fill=none,color=black,scale=0.8] at (-0.2,0.3) {$0.4$};
  \node[fill=none,color=black,scale=0.8] at (-0.5,0.7) {$0.4$};
  \node[fill=none,color=black,scale=0.8] at (0.5,0.3) {$0.4$};
  \node[fill=none,color=black,scale=0.8] at (0.2,1.5) {$0.4$};
  \node[fill=none,color=black,scale=0.8] at (-0.7,1.5) {$0.7$};

  \draw[->-,color=black] (new) .. controls (-1.1,3.5) .. (i);
  \node[fill=none,color=black,scale=0.8] at (-1.2,3.6) {$0.5$};
  \node[fill=none,color=black,scale=0.8] at (-1.5,0.6) {$0.5$};

\end{tikzpicture}}}
\caption{An example of a portion of the WDG $D$  before  and after  switching along path $j\rightarrow i \rightarrow j'$.
This figure illustrates the case when $|w(i,j')|<|w(j,i)|$, hence $P =|w(i,j')|$. 
}\label{switchingIl1} 
\end{figure}

\begin{figure}[!h]
\centering
%
{
\subfloat{
\begin{tikzpicture}[scale=1.3] 
  \node[fill=none, scale=0.9, color=black]  () at (0,-1) {(a) Before switching}; 
  \node[draw, circle, scale=0.7] (a) at (-1,0) {}; 
  \node[draw, circle, scale=0.7] (b) at (0,0) {}; 
  \node[draw, circle, scale=0.7] (c) at (1,0) {}; 
  \node[draw, circle, scale=0.7] (f) at (0,2) {}; 
  \node[draw, circle, scale=0.7] (g) at (-1,2) {};
  \node[draw, circle, scale=0.7] (i) at (0,4) {}; 
  \node[draw, rectangle,fill=black, scale=0.8] (d) at (0,1) {}; 
  \node[draw, rectangle,fill=black, scale=0.8] (e) at (-1,1) {}; 
  \node[draw, rectangle,fill=black, scale=0.8] (h) at (0,3) {};

  \node[fill=none,color=black,scale=0.8] at (-0.3,1) {$j$};
  \node[fill=none,color=black,scale=.8] at (-0.3,2) {$i$};
  \node[fill=none,color=black,scale=.8] at (-0.3,3) {$j'$};
  \node[fill=none,color=black,scale=.8] at (-0.3,4) {$i'$};
  \node[fill=none,color=black,scale=.8]  at (0.7,2) {$\gamma_i=+1$};
  \node[fill=none,color=black,scale=.8] at (1.3,0) {$k$};

  \draw[->-,color=black] (a) --  (d);
  \draw[->-,color=black] (b) --  (d);
  \draw[->-,color=black] (c) --  (d);
  \draw[->-,color=red] (d) -- (f);
  \draw[->-,color=black] (e) -- (f);
  \draw[->-,color=red] (f) -- node[auto]{$\scriptstyle 0.9$}(h);
  \draw[->-,color=black] (g) -- node[auto]{$\scriptstyle 0.9$}(h);
  \draw[->-,color=black] (c).. controls (2,2) ..     (h);
  \draw[->-,color=black] (h) -- node[auto]{$\scriptstyle 0.9$} (i);
  \node[fill=none,color=black,scale=0.8] at (1.4,1.5) {$0.9$};
  \node[fill=none,color=black,scale=0.8] at (-0.2,0.3) {$0.5$};
  \node[fill=none,color=black,scale=0.8] at (-0.5,0.7) {$0.5$};
  \node[fill=none,color=black,scale=0.8] at (0.5,0.3) {$0.5$};
  \node[fill=none,color=red,scale=0.8] at (0.2,1.5) {$0.5$};
  \node[fill=none,color=black,scale=0.8] at (-0.7,1.5) {$0.3$};
\end{tikzpicture}}}
\qquad
{
\subfloat{
\begin{tikzpicture}[scale=1.3]
 \node[fill=none, scale=0.9, color=black]  () at (0,-1) {(b) After switching}; 
  \node[draw, circle, scale=0.7] (a) at (-1,0) {}; 
  \node[draw, circle, scale=0.7] (b) at (0,0) {}; 
  \node[draw, circle, scale=0.7] (c) at (1,0) {}; 
  \node[draw, circle, scale=0.7] (f) at (0,2) {}; 
  \node[draw, circle, scale=0.7] (g) at (-1,2) {};
  \node[draw, circle, scale=0.7] (i) at (0,4) {}; 
  \node[draw, rectangle,fill=black, scale=0.8] (d) at (0,1) {}; 
  \node[draw, rectangle,fill=black, scale=0.8] (e) at (-1,1) {}; 
  \node[draw, rectangle,fill=black, scale=0.8] (h) at (0,3) {};
  \node[draw, rectangle, fill=black,scale=0.7] (new) at (-2,2) {};

  \node[fill=none,color=black,scale=0.8] at (-0.3,1) {$j$};
  \node[fill=none,color=black,scale=.8] at (-0.3,2) {$i$};
  \node[fill=none,color=black,scale=.8] at (-0.3,3) {$j'$};
  \node[fill=none,color=black,scale=.8] at (-0.3,4) {$i'$};
  \node[fill=none,color=black,scale=.8]  at (0.7,2) {$\gamma_i=+1$};
  \node[fill=none,color=black,scale=0.8] at (1.4,1.5) {$0.4$};
  \draw[->-,color=black] (c).. controls (2,2) ..     (h);
  \node[fill=none,color=black,scale=.8] at (-2.3,2) {$j''$};
  \node[fill=none,color=black,scale=.8] at (1.3,0) {$k$};

  \draw[->-,color=black] (e) -- (f);
  \draw[->-,color=black] (f) -- node[auto]{$\scriptstyle 0.4$}(h);
  \draw[->-,color=black] (g) -- node[auto]{$\scriptstyle 0.4$}(h);
  \draw[->-,color=black] (h) -- node[auto]{$\scriptstyle 0.4$} (i);
  \node[fill=none,color=black,scale=0.8] at (-0.7,1.5) {$0.3$};
  \draw[->-,color=black] (a) -- (new);

  \draw[->-,color=black] (b) .. controls (-1,-1) and (-2:-2) .. (new);
  \node[fill=none,color=black,scale=0.8] at (-1.8,0) {$0.5$};
  \draw[->-,color=black] (g) --   node[auto]{$\scriptstyle 0.5$}(new);
  \draw[->-,color=black] (new) .. controls (-1.1,3.5) .. (i);
  \node[fill=none,color=black,scale=0.8] at (-1.2,3.6) {$0.5$};
  \node[fill=none,color=black,scale=0.8] at (-1.5,0.6) {$0.5$};

\end{tikzpicture}}}
\caption{An example of a portion of the WDG $D$  before and after   switching along path $j\rightarrow i \rightarrow j'$.
This figure illustrates the case when $|w(i,j')|>|w(j,i)|$, hence $P =|w(j,i)|$. 
}\label{switchingIl2} 
\end{figure}
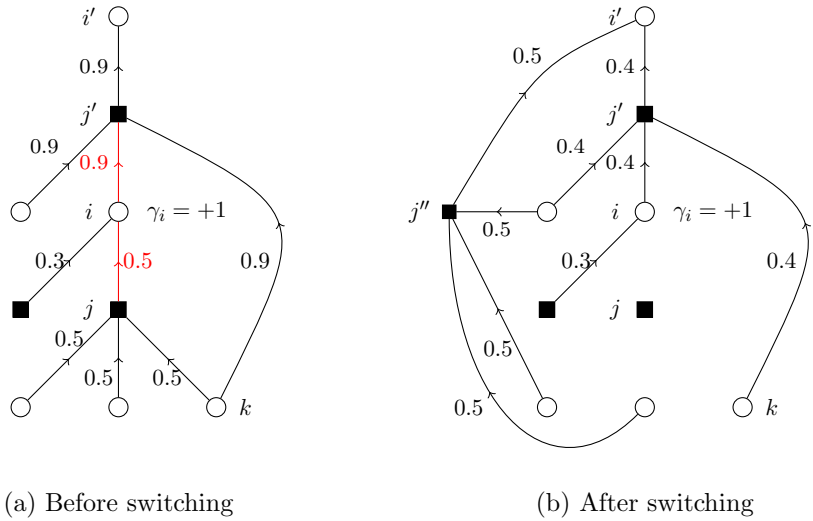

\begin{claim}[Switch operation properties] \label{switchingLemma}
Let $i\in V$ and $j,j'\in C$ such 
that $(j,i)$ and  $(i,j')$ are edges in $D$. After switching $D$ along  $j\rightarrow i \rightarrow j'$,  the followings hold:  
\begin{itemize}
\item[a)]  $D$ is still an  acyclic WDG of a hyperflow for $\gamma$ in  $\overline{G}$.  
\item[b)]  For each  variable node $v\in V$, the total inflow  $F_v^{in}(w)$
to $v$ and the total outflow $F_v^{out}(w)$
from $v$ do not increase.
\item[c)] The indegree of $i$  and the outdegree  of $i$  do not increase and at least one of them  decreases by at least one. 
\end{itemize}
\end{claim}
{\em Proof.}   
First we note that due to the acyclicity of $D$, variable node $i'$ will not cancel out after XORing $j$ and $j'$ in Line \ref{L5}. 
Indeed, assume that $i'$ cancels out, then $i'$ must be  connected to $j$ (by an edge 
incoming from $i'$ since $j$ already has an edge outgoing to $i$), hence we get the  cycle  $j\rightarrow i\rightarrow j' \rightarrow i' \rightarrow j$.  

It is straightforward
 to verify  (b) and  (c).  Note that the only variable nodes in $D$ whose inflow or outflow change are those shared by $j$ and $j'$ -- namely, $i$ 
 and possibly other nodes  $k$ (see  Figures \ref{switchingIl1}  and \ref{switchingIl2}).  
Both the inflow to $i$ and the outflow from $i$  decrease by $P$, the  outflow from $k$ decreases by $P$  and the inflow to $k$ remains unchanged. 

It is also straightforward to verify   that the acyclicity of $D$ 
and the  WDG variable nodes inequalities ((d) in Definition \ref{dualdefs}) 
are maintained.
To complete the proof of (a), we need to show that  the hyperflow check  nodes  inequalities ((c) in Definition \ref{dualdefs}) are maintained. In particular, we have to argue that in Line \ref{L6} 
it is not possible that check node $j''$ is already present  
with a different edge orientation, i.e., with 
an edge outgoing from $j''$ to a variable node $i''\neq i'$. Again, this follows from the acyclicity of $D$.  
Assume that right before executing Line \ref{L6}, 
 there is an  edge outgoing from $j''$ to a variable node $i''\neq i'$. 
Since variable $i''$ appears in check $j''$, then 
it appears in either $j$ or $j'$,  hence   either $(i'',j)$  or $(i'',j')$ is an edge in $D$.
If $(i'',j)$ is an edge,  we get the cycle  $i''\rightarrow j\rightarrow i \rightarrow j' \rightarrow i' \rightarrow j'' \rightarrow i''$.
If $(i'',j')$ is an edge,  we get the cycle  $i''\rightarrow j'\rightarrow  i'\rightarrow j'' \rightarrow i''$.
\lemfinito

Algorithm \ref{primitivization} given below iteratively modifies  $D$ until it  becomes   primitive by repeated application of the 
Switch operation. Recall that $In_D(i)$ is the  set of check nodes incident to edges ingoing to $i$ and 
$Out_D(i)$ is the  set of check nodes incident to  edges outgoing from $i$.

\begin{algorithm}
\caption{Making  the WDG $D$ primitive} \label{primitivization}
\begin{algorithmic}[1]
\For{each variable node $i\in V$} 
\While{$InDegree_D(i)\neq 0$ and $OutDegree_D(i)\neq 0$ (i.e., $In_D(i)\neq \emptyset$ and  $Out_D(i)\neq \emptyset$)}  
\State Pick any  $j\in In_D(i)$ and any $j'\in Out_D(i)$
\State Switch $D$ along $j\rightarrow i\rightarrow j'$ 
\EndWhile
\EndFor
\For{each variable node $i\in V$ such that $\gamma_i>0$ and $InDegree_D(i)\neq 0$} 
\State  Remove all the  edges in $D$ connected to check nodes in $In_D(i)$
\EndFor
\end{algorithmic}
\end{algorithm}

For each $i\in V$, Part (c) of Claim \ref{switchingLemma} asserts that 
the indegree and the outdegree of $i$  
do not increase and at least one of them 
decreases by at least one, hence the inner while-loop halts in a finite number of steps.
Thus at the end of each iteration of the first outer for-loop, variable node $i$ has 
either zero indegree or  zero outdegree. 
 Part (b) of Claim 
\ref{switchingLemma} guarantees that once the  indegree or the outdegree of a node $i$ is zero, 
it remains zero in future iterations of the algorithm.

Consider $D$ after  the end of the first outer for-loop and consider any variable node $i\in V$.

If $\gamma_i\leq 0$, the indegree of $i$ must be nonzero due to the  WDG variable nodes  inequalities. 
Thus the outdegree of $i$ must be zero.

If $\gamma_i> 0$ and the indegree  of $i$ is nonzero, then the outdegree of $i$ must be zero, hence 
the outflow from $i$ is zero.
Since $\gamma_i>0$ and the outflow from $i$ is zero, the inflow to $i$ is unnecessary. 
The second for-loop performs a final pass to removes this unnecessary inflow by 
 disconnecting the edges of the check nodes in $In_D(i)$ from $D$ (thus now both the  
indegree and the outdegree of $i$ are zeros).

\finito

\section{Impact of redundant checks}\label{impred}

In this section we establish Theorem \ref{crr1} and Corollary  \ref{crr2.2} restated below for convenience. 
The proof of Theorem \ref{crr1}  uses the LP excess lemma \cite{BGU14}. 

\begin{lemma}[\cite{BGU14}]  \label{excesslem} {\em\bf (LP Excess Lemma: trading crossover probability  with LP excess)} 
Let $H = (V,C,E)$ be a Tanner graph.  
Let $0<\e<\e'<1$ and $0< \delta< 1$ such that $\epsilon' = \epsilon + (1-\epsilon)\delta$. 
Let $q_{\epsilon'}$ be the probability  that the LP decoder of $H$  fails on the $\epsilon'$-BSC. 
Consider operating on the $\e$-BSC, i.e.,  choose the error pattern $x \sim \operatorname{Ber}(\epsilon,n)$.
Then the probability  that there exists a dual witness in $H$ for $(-1)^x-\frac{\delta}{2}$ is at least $1-\frac{2q_{\epsilon'}}{\delta}$.

In other words, if we let $(-1)^{x_i}-  \sum_{j \in N(i)} w(i,j)$ be the ``LP excess'' of $w$ on variable node $i$, 
then the probability over the $\epsilon$-BSC that there exists a dual witness with LP excess greater than $\frac{\delta}{2}$ on all the variable nodes is at least $1-\frac{2q_{\epsilon'}}{\delta}$.
\end{lemma}

\noindent 
{\bf Theorem \ref{crr1}}~{\bf  (High degree redundant checks  do not improve LP threshold)} 
{\em 
Let $\G = \{G_n\}_n$ be an  infinite   family of  Tanner graphs such that each check node has degree at most $d$, where $d$ is a constant.  
Assume that  $\G$ is asymptotically strong. 
 Then:
\begin{itemize}
\item[a)] For any small constant  $\delta>0$, there exists a sufficiently large constant $k\geq d$ 
$($dependent on $\delta$ and 
independent of $n$$)$
  such that 
       $\xi_{LP}(\overline{\G}^k)\geq  \xi_{LP}(\overline{\G}) - \delta$.   
\item[b)] If  $k(n)$  is a real valued function of $n$ such that $k(n) = \omega(1)$ $($i.e., $k(n)$ tends to infinity as $n$ tends to infinity$)$, 
then $\xi_{LP}(\overline{\G}^{k})=  \xi_{LP}(\overline{\G})$.
\end{itemize}
}
~\\
\noindent 
{\bf Proof:} Part (b) is an immediate consequence of (a). 
At a high level, the argument behind (a) is as follows. 
We will  operate $\overline{\G}$ on the BSC slightly below its LP threshold to guarantee the existence of a dual witness $w$  with some small but constant 
LP excess over all variable nodes. Namely, we set the LP excess to $\frac{\delta}{4}$. 
Since $\overline{\G}$  contains all redundant check nodes, we can assume that $w$ is {\em primitive}. 
 We will trim $w$ by removing  all check nodes of degree larger than $k$. The trimming process  leads  to a distorted dual witness $w^k$, 
where the   variable nodes inequalities are violated for $w^k$  over some set of variables which we call {\em problematic}.   
Call a variable {\em risky} if it receives 
at least $\frac{\delta}{8}$  flow from the removed check nodes and let $U$ be the set of  risky variables. 
Thus the  risky  variables include all the problematic variables. Moreover, all the risky variables are received in error since $w$ is primitive. 
Due to the high degree of the removed check nodes  and due to the primitivity of $w$,  the removed checks give the 
variables in error   little flow, namely at most $\frac{n}{k-1}$.   
It follows that  the set $U$ of risky variables  is small, namely $|U|\leq \frac{8n}{\delta(k-1)}$. 
Due to the  primitivity of $w$,
 the variables in error, and in particular the 
problematic variables, have no outgoing edges. That is, the outflow from each  problematic variable node is zero, 
hence fixing each  problematic variable   requires adding a unit flow in the worst case (this conclusion critically depends on the primitivity of $w$).  
By construction, the nonrisky variables  still have $\frac{\delta}{4}-\frac{\delta}{8}=\frac{\delta}{8}$ LP excess after the trimming process.
We will use this remaining excess to fix $w^k$ by patching a dual witness which turns the remaining small LP excess on the nonrisky 
variables  into a unit flow on each risky variable.  
The existence of the patch follows from the {\em asymptotic strength} of $\G$. 
 
More formally, 
let $\delta>0$ and assume without loss of generality that $\xi_{LP}(\overline{\G})>0$ and   $\delta < \xi_{LP}(\overline{\G})$ 
 (otherwise, the claim of the theorem is trivial).  
We will show that there is a sufficiently large constant  $k$ such that       $\xi_{LP}(\overline{\G}^k)\geq \xi_{LP}(\overline{\G})-\delta$. 
Let  $\e = \xi_{LP}(\overline{\G})-\delta$ and $\e' = \e+(1-\e)\frac{\delta}{2}$, thus $0<\e<\e'< \xi_{LP}(\overline{\G})$. 
 Let $q_{\e'}(n)$ be the probability of error of the LP decoder of $\overline{G_n}$ 
over the $\e'$-BSC.  Note that $q_{\e'}(n)$ tends to zero as $n$ tends to infinity since  $\e'< \xi_{LP}(\overline{\G})$.  
By the LP excess lemma (Lemma \ref{excesslem}), with probability  at least $1-\frac{4q_{\epsilon'}(n)}{\delta}$, 
there exists a dual witness in $\overline{G_n}$ for $(-1)^x-\frac{\delta}{4}$, where   $x \sim \operatorname{Ber}(\epsilon,n)$. 
In what follows, consider any  $k$ and $n$ such that $d \leq k \leq n$,  consider any $x\in \{0,1\}^n$ such that $(-1)^x-\frac{\delta}{4}$ has 
a dual witness $w$ in $\overline{G_n}$, say that  $\overline{G_n} = (V,\overline{C}, \overline{E})$  and 
consider the Tanner graph $\overline{G_n}^k = (V,\overline{C}^k, \overline{E}^k)$.   
We will construct from $w$ a dual witness for $(-1)^x$ in $\overline{G_n}^k$  for sufficiently large $k$.

 Let  $V^+_x = \{ i \in V : (-1)^{x_i}-\frac{\delta}{4}\geq 0\}$ and  $V^-_x = \{ i \in V : (-1)^{x_i}-\frac{\delta}{4}< 0\}$.
Note that since $0<\delta<1$,  $V^+_x = \{ i \in V : (-1)^{x_i}=1\}$ and   
$V^-_x = \{ i \in V : (-1)^{x_i} =-1\}$, i.e.,   $V_x^+$ is the set of variable nodes received correctly 
and $V_x^-$  consists of those received in error. 
Since $\overline{G_n}$ contains all redundant check nodes, we can assume by   Lemma \ref{primlem} that the WDG $D$ of $w$ is 
a {\em primitive hyperflow}. 
Since $D$ is  a primitive hyperflow, for each check node $j$ in $D$,  all the ingoing edges to $j$ are 
from variables in $V^+_x$  and the only outgoing edge from $j$ is to some variable  in $V^-_x$. 
Let $L^k$ be the set of check nodes in $\overline{G_n}$ of degree larger than $k$, i.e., $L^k = \overline{C}-\overline{C^k}$. 
The check nodes in $L^k$ give the variable nodes in $V^-_x$ a total flow which is at most 
$\frac{|V^{+}_x|}{k-1}  \leq \frac{n}{k-1}$.  
Call a variable node in  $V^-_x$ {\em risky} if it receives at least $\frac{\delta}{8}$  flow in total from the checks in  $L^k$. 
Let $U$ be the set of risky variable nodes, 
thus $$|U|\leq \frac{8n}{\delta (k-1)}.$$   
Remove from $D$ all the check nodes in $L^k$ and all the associated  edges 
and let $w^k$ be the resulting  weight map $w^k: \overline{E}^k \rightarrow \R$. 
The map $w^k$ possibly violates the variable nodes inequalities over some variables in $U$, 
but it satisfies the hyperflow check nodes inequalities and hence the dual witness 
check nodes inequalities over all checks.
 For each $i\in V$, consider the flows at $i$ associated with $w$ and $w^k$: 
$F_{i}(w) = \sum_{j} w(i,j)$,
$F_{i}^{out}(w) = \sum_{j\leftarrow  i} |w(i,j)|$,
$F_{i}(w^k) = \sum_{j} w^k(i,j)$,
$F_{i}^{in}(w^k) = \sum_{j\rightarrow i} |w^k(i,j)|$ and 
$F_{i}^{out}(w^k) = \sum_{j\leftarrow  i} |w^k(i,j)|$.
Since $w$ is primitive,  
none of the variables $i\in V_x^-$ have  
outgoing edges in $\overline{G}^k$, thus $F_{i}^{out}(w)= 0$ and hence 
$F_{i}^{out}(w^k)= 0$. 
Thus for each $i\in U$,   $F_{i}(w^k) = - F_{i}^{in}(w^k)  \leq 0$.  
If $i\in V_x^{-} - U$, we have
\[
F_{i}(w^k) <  F_i(w) + \frac{\delta}{8} <  -1-\frac{\delta}{4} +\frac{\delta}{8} = -1-\frac{\delta}{8}.
\]  
If $i\in V_x^{+}$, we have 
$F_i(w^k) \leq  F_i(w)  <  1-\frac{\delta}{4}<  1-\frac{\delta}{8}$.  
Therefore,  for each variable $i\in V$, 
\begin{equation}\label{lkjdsa1}
\left\{\begin{array}{llll} 
F_i(w^k) &\leq & 0 & \mbox{ if } i\in U \\
F_i(w^k) &<& (-1)^{x_i}-\frac{\delta}{8}  & \mbox{ otherwise. } 
\end{array}\right.
\end{equation}
To turn $w^k$ into a dual witness for $(-1)^x$, we have to fix the possible violations 
of  variable nodes inequalities over $U$. Over $V-U$, 
the variable nodes inequalities are satisfied with $\frac{\delta}{8}$ excess.  
We will use this excess to fix the problematic variables in  $U$ by patching to $w^k$ a dual witness for the 
asymmetric LLR vector $\gamma\in \R^V$ given by 
\[                        
\gamma_i = \left\{\begin{array}{ll} -1 &\mbox{ if } i\in U  \\ \frac{\delta}{8} & \mbox{ otherwise, } \end{array}\right.
\]
for all $i\in V$.

Since $\G$ is {\em asymptotically strong}, there exists a constant 
$\alpha_{\delta} >0$ (independent of $n$) such that if $|U|\leq \alpha_\delta n$, 
 the LP decoder of $G_n = (V,C,E)$ succeeds on the asymmetric LLR vector $\gamma$.  
Hence, if  $\frac{8}{\delta (k-1)} \leq \alpha_\delta$, then $\gamma$ has a dual witness 
$v: E \rightarrow \R$ in $G_n$.  Since $k \geq d$ (recall that $d$ is the maximum degree of a check node in $\G$), we can extend $v$ from $E$ to $\overline{E}^k$ by zeros. 
Let   $v^k: \overline{E}^k \rightarrow \R$
 be the resulting weight map, thus  
\begin{equation}\label{lkjdsa2}
\left\{\begin{array}{llll} 
    F_i(v^k) &<& -1 & \mbox{ if } i\in U \\
F_i(v^k) &<& \frac{\delta}{8} & \mbox{ otherwise, } 
\end{array}\right.
\end{equation}
where    $F_i(v^k) = \sum_{j} v^k(i,j)$. Since $U \subset V_x^-$, it follows from (\ref{lkjdsa1}) and (\ref{lkjdsa2}) that 
$F_i(w^k)+F_i(v^k)< (-1)^{x_i}$, for all $i\in V$.  
Noting that the dual witness  check nodes inequalities are preserved by   superposition,  we conclude that 
 $w^k + v^k$ is the desired dual witness of $(-1)^x$.

In summary,
for  all   $\delta >0$ such that $\delta< \xi_{LP}(\overline{\G})$, there exists 
a constant $\alpha_\delta> 0$ such that with  $\e = \xi_{LP}(\overline{\G})-\delta$, $\e' = \e+(1-\e)\frac{\delta}{2}$ and 
$k = \lceil \frac{8}{\delta\alpha_\delta} \rceil +1$,  the following holds for all values of $n$.  
 Let $q_{\e'}(n)$ be the probability of error of the LP decoder of $\overline{G}_n$ over the $\e'$-BSC. 
Then there exists a dual witness in $\overline{G_n}^k$ for $(-1)^x$ 
 with probability  at least $1-\frac{4q_{\epsilon'}(n)}{\delta}$ over the choice of  
  $x \sim \operatorname{Ber}(\epsilon,n)$. 
Since $\e'< \xi_{LP}(\overline{\G})$, $q_{\e'}(n)$ tends to zero as $n$ increases.
It follows that, for all $\delta>0$, there exists a sufficiently large constant $k>0$ 
dependent on $\delta$ such that $\xi_{LP}(\overline{\G}^k)\geq  \xi_{LP}(\overline{\G}) - \delta$. 
 \finito

To derive Corollary \ref{crr2.2} from  Theorem \ref{crr1}, we need the following classical result.
\begin{theorem}[\cite{Fel03}]\label{treesp} {\em\bf (Optimality of LP decoding on 
acylic graphs)} 
Let $H = (V,C,E)$ be a Tanner graph and  $Q_H$ the associated code. If  $H$ is acyclic, 
then the fundamental polytope $P(H)$ of $H$ is the convex span of the code $Q_H$, 
i.e.,   $conv(Q_H) = P(H)$.  
\end{theorem}
See \cite{KVLarge} or \cite{BG11} for a proof. 
It follows from Theorem \ref{treesp}  that redundant checks obtained by acyclic sums do not tighten the polytope.
A statement similar to Corollary \ref{treespc} appears in \cite{BG11}. We include a short derivation of Corollary \ref{treespc} 
from Theorem \ref{treesp} for completeness.
\begin{corollary}[Acyclic  redundnat checks do not tighten the polytope]\label{treespc}
Let $G = (V,C,E)$ be a Tanner graph and $Q \subset \F_2^n$ the associated code. 
For each check $j \in C$, let  $z_j\in Q^\bot$ be   the vector in the dual code associated with $j$. 
Let $D\subset Q^\bot$ such that each check $z\in D$ is obtained by an acyclic sum of checks of $G$. 
That is, each $z\in D$ is of the form $z = \sum_{j \in S} z_j$,  for some $S\subset C$  such that  
the graph induced by $G$ on $S$ is acyclic. 
Consider the Tanner graph  $G' = (V,C\cup D, E')$ resulting from $G$ by adding 
all the checks in $D$. Then $P(G) = P(G')$. 
\end{corollary}
{\bf Proof:} By definition, 
$P(G) = \bigcap_{j \in C} \operatorname{conv}(Q_j)$ and 
$P(G') =\bigcap_{z \in C\cup D} \operatorname{conv}(Q_z)$. 
Consider any check $z\in D$. It is enough  to argue that $P(G)\subset conv(Q_z)$. 
Let $S\subset C$  such that  $z = \sum_{j \in S} z_j$ and  
the graph  $G_S = (V_S,S,E_S)$  induced by $G$ on $S$ is acyclic.  
By Theorem \ref{treesp}, $P(G_S)=conv(Q_{G_S})$.  
Extending the polytopes from $\R^{V_S}$ to $\R^{V}$, we get $\bigcap_{j \in S} \operatorname{conv}(Q_j)=conv(Q^S)$,   
where $Q^S$ is the supercode of $Q$ consisting of all the vectors in $\F_2^n$ satisfying all the checks in $S$.  
Since $z$ is a linear combinations of checks  in $S$, we have 
$Q^S \subset Q_z$, hence $conv(Q^S) \subset conv(Q_z)$.  Therefore  
\[
P(G)  = \underset{j \in C} \bigcap \operatorname{conv}(Q_j) \subset \underset{j \in S} \bigcap \operatorname{conv}(Q_j) =conv(Q^S)\subset conv(Q_z). 
\] 
\finito

Finally, we conclude Corollary \ref{crr2.2} from Theorem \ref{crr1} and   Corollary \ref{treespc}.\\

\noindent 
{\bf Corollary \ref{crr2.2}}~{\bf  (Redundant checks do not improve LP threshold)} 
{\em 
Let $\G=\{G_n\}_n$ be an  infinite   family of Tanner graphs of bounded check degree. 
If $\G$ is asymptotically strong and rigid, then $\xi_{LP}(\overline{\G}) = \xi_{LP}(\G)$.  
}\\

\noindent 
{\bf Proof:}   
Say that $G_n = (V_n,C_n,E_n)$ and $\overline{G_n} = (V_n,\overline{C_n},\overline{E_n})$. 
Since $\G$ is rigid, $\Delta(G_n) = \omega(1)$.   
Let $k(n):=\Delta(G_n)-1$ and assume that $n$ is large enough so that $k(n)$ is at least 
the maximum  degree of a check node of $\G$.
By the definition of $k(n)$,  all redundant checks in $\overline{C_n}$
of degree at most $k(n)$ 
are obtained  by  acyclic sums of checks in $C_n$. 
By  Corollary \ref{treespc}, 
$P(G_n) = P(\overline{G_n}^{k(n)})$, hence  $ \xi_{LP}(\G)=\xi_{LP}(\overline{\G}^k)$.
On the other hand, by Theorem \ref{crr1}, 
$\xi_{LP}(\overline{\G}^k) = \xi_{LP}(\overline{\G})$ since $k(n) = \omega(1)$ and $\G$ is asymptotically strong. 
 It follows that $\xi_{LP}(\G)=\xi_{LP}(\overline{\G})$.  
\finito

\section{Expansion and asymptotic strength} \label{expanstr}

In this section we prove Theorem \ref{expstr} restated below for convenience. The proof uses the notion of 
a narrow dual witness defined below. 

\begin{definition}[Narrow dual witness]
Let $G=(V,C,E)$ be a Tanner graph, $y\in \{0,1\}^n$ an error vector and $w:E \rightarrow \R$ a 
 dual witness for $(-1)^y$ in $G$. We call $w$ a {\em narrow} dual witness for $(-1)^y$ if 
 all the edges not incident to $N(U)$ have zero weights, 
where $U = \{ i\in V : y_i =1\}$ is the set of variables in error 
$($i.e, if an edge is not incident to a check node incident to a variable in error, then it has zero weight$)$.
\end{definition}
A key property of a narrow dual witness is that the flow at the  correct variable  nodes 
far from $U$ by more than $2$ edges is zero.

Recall that a Tanner graph $G=(V,C,E)$ is called an {\em $(\varepsilon n, \K)$-expander} if for each subset $S \subset V$ of variable nodes of size at most 
$\varepsilon n$, we have $|N(S)| \geq \K |S|$. 

Feldman at al. \cite{FMS07} argued that 
the LP decoder of graphs with good expansion corrects a positive fraction of errors. Although not explicitly stated, the dual witness constructed in their proof 
is actually narrow.  Their argument was later simplified by Viderman \cite{Vid13} who also improved the expansion requirement. 

\begin{lemma}[Implicit in \cite{Vid13}]  \label{fmsflow}    
{\em \bf (Expansion implies  the existence of a narrow dual witness)} 
Let $d_v>0,\varepsilon>0$ and  $\delta > \frac{2}{3}$ be constants such that $d_v$ is an integer and 
  $\delta d_v$ is an integer.
Let $G=(V,C,E)$ be a Tanner graph with regular variable degree  $d_v$ and assume that $G$ is an  $(\varepsilon n, \delta d_v)$-expander.  
Then $(-1)^y$ has a {\em narrow} dual witness in $G$, for each error vector  $y\in \{0,1\}^n$ of weight at most 
$\frac{3\delta -2}{2\delta -1}(\varepsilon n-1)$. 
\end{lemma}

\noindent 
{\bf Theorem \ref{expstr}}~{\bf  (Expansion implies  asymptotic strength)} 
{\em
Let $d_v>0,\varepsilon>0$ and  $\delta > \frac{2}{3}$ be constants such that $d_v$ is an integer and 
  $\delta d_v$ is an integer. 
Let $\G = \{G_n\}_n$ be an  infinite   family of   Tanner graphs with regular variable degree $d_v$  and bounded check degree. 
If $G_n$ is an  $(\varepsilon n, \delta d_v)$-expander for each $n$,   
then $\G$ is asymptotically strong. 
}

~\\
\noindent 
{\bf Proof:}    
The proof is based on successive superpositions of narrow dual witnesses obtained from 
Lemma \ref{fmsflow}  to amplify the flow at the variable nodes in errors. 
 The fact they are narrow is essential for superposing them 
without violating the variable nodes constraints at the correct variables.

Consider any constant  $\beta>0$ and let $B = \lceil\frac{1}{\beta}\rceil$. 
It is enough to find a constant 
$\alpha>0$ and construct, for each $n$ and each   $U\subset V = \{1,\ldots,n\}$ of size most $\alpha n$, 
 a dual witness $w$ in $G_n=(V,C,E)$ for the asymmetric LLR vector $\gamma\in \R^V$ given by 
\[                        
\gamma_i = \left\{\begin{array}{ll} -B &\mbox{ if } i\in U  \\ 1 & \mbox{ otherwise,}\end{array}\right.
\]
for all $i\in V$. Since $B \geq \frac{1}{\beta}$,  the scaled version $\frac{1}{B}w$ of $w$ is the desired dual witness for $\gamma({y,\beta})$ 
(as given in Definition \ref{asstrong}), where $y\in \{0,1\}^n$ is the indicator vector of $U$.

If $S\subset V$ is a set of variable nodes and $t\geq 0$ is an integer, let $N_{var}(S; t)$ be the set of variable nodes  at distance at most $2t$ from $S$.
Thus $N_{var}(S; 0)=S$ and $N_{var}(S; 1)$ is the set of variables connected to check nodes connected to  $S$.

Let $\alpha >0$ be a sufficiently small constant such that for each $U\subset V$ of size at most $\alpha n$, we have 
\begin{equation}\label{condsuc}
|N_{var}(U;B-1)|\leq \frac{3\delta -2}{2\delta -1}(\varepsilon n-1), 
\end{equation}
for sufficiently  large $n$ (the explicit value of  $\alpha$ is  at the end of the proof). Assume that $|U|\leq \alpha n$ and let $U^t = N_{var}(U; t)$, for $t = 0,\ldots, B-1$. 
In what follows,  consider any $t \in\{ 0,\ldots, B-1\}$. 
Since $|U^t|\leq \frac{3\delta -2}{2\delta -1}(\varepsilon n-1)$,  
Lemma \ref{fmsflow} guarantees that 
 $(-1)^{y^t}$ has a narrow dual witness $w^t: E\rightarrow \R$ in $G$, 
where $y^t\in \{0,1\}^n$ is the indicator vector of $U_t$, i.e., $y_i^t=1$ iff $i\in U^t$. 
The fact that 
 $w^t$ is narrow  means all the edges not incident to $N(U^t)$ have zero weights, thus the 
flow at the variable nodes outside $U^{t+1}$ is zero.  That is, 
 $F_i(w^t) =  0$ for each $i\in V - U^{t+1}$, where 
 $F_i(w^t) = \sum_{j} w^t(i,j)$ is the flow with respect to $w^t$ at  variable node $i$.
 Let  $w = \sum_{t=0}^{B-1} w^t$. 
We will argue that $w$ is the desired  dual witness for $\gamma$.

First, note that   superposing dual witnesses does not  violate the 
dual witness check nodes inequalities ((b) in Definition \ref{dualdefs}).
 Thus, we only have to worry about the variable nodes inequalities ((a) in Definition \ref{dualdefs}).
Consider the flow at the variable nodes  with respect to $w$: $F_i(w) = \sum_{t=0}^{B-1}F_i(w^t)$, for all $i \in V$.
  We have to show that
\begin{equation}\label{mib2}
\left\{\begin{array}{llll}
F_i(w) &<& -B &\mbox{ if } i\in U^0 =U \\ 
F_i(w) &<& 1  &\mbox{ otherwise.}
\end{array}\right.
\end{equation}                        
Since 
each $w^t$ is a narrow dual witness for $(-1)^{y^t}$, we have 
\[                        
\left\{\begin{array}{llll} F_i({w^t})&<& -1 &\mbox{ if } i\in U^t  \\ 
                         F_i({w^t})&<& 1 &\mbox{ if } i\in  U^{t+1}-U^t  \\ 
                         F_i({w^t})&=& 0 & \mbox{ if } i\in V - U^{t+1}.
\end{array}\right.
\]
Summing over $t=0,\ldots, B-1$ and using the fact that 
 $U^0\subset U^1 \subset  U^2\subset  \ldots \subset U^{B}$, we obtain 
\[                        
\left\{\begin{array}{llll}
F_i(w) &<& -B &\mbox{ if } i\in U^0  \\ 
F_i(w) &<& -(B-2) &\mbox{ if } i\in U^1-U^0 \\ 
F_i(w) &<& -(B-3) &\mbox{ if } i\in U^2-U^1 \\ 
F_i(w) &<& -(B-4) &\mbox{ if } i\in U^3-U^2 \\ 
\ldots & \\\
F_i(w) &<& -2 &\mbox{ if } i\in U^{B-3}-U^{B-4} \\                       
F_i(w) &<& -1 &\mbox{ if } i\in U^{B-2}-U^{B-3} \\                       
F_i(w) &<& 0 &\mbox{ if } i\in U^{B-1}-U^{B-2} \\                             
F_i(w) &<& 1 &\mbox{ if } i\in U^{B}-U^{B-1} \\                          
F_i(w)&=& 0 &\mbox{ if } i\in V-U^{B},
\end{array}\right.
\]
and hence (\ref{mib2}) follows.

Finally, note that if  $d_c$ be the maximum check degree of check node in $G_n$ for all $n$, then for all $t\geq 0$, 
\[
|N_{var}(U;t)|\leq \sum_{i=0}^{t} (d_v(d_c-1))^i|U| = \frac{(d_v(d_c-1))^{t+1}-1}{d_v(d_c-1)-1}|U|.
\]
Thus condition (\ref{condsuc}) is satisfied if  
\[
  \frac{(d_v(d_c-1))^{B}-1}{d_v(d_c-1)-1}\alpha n \leq  \frac{3\delta -2}{2\delta -1}(\varepsilon n-1),
\]
which holds for $n$ sufficiently large  with 
\[
\alpha =  \frac{(3\delta -2)(d_v(d_c-1)-1) }{(4\delta -2)((d_v(d_c-1))^{\lceil\frac{1}{\beta}\rceil}-1)} \varepsilon. 
\]

\finito

\section{Nondegeneracy of random graphs}\label{randSec}

In this section we prove Lemmas \ref{lcrr3} and \ref{randlem} restated below for convenience.\\

\noindent
{\bf Lemma \ref{lcrr3}}~{\bf  (Rigidity versus girth and nondegeneracy)} 
{\em
Let $\G=\{G_n\}_n$ be an  infinite   family of Tanner graphs of bounded check degree.  
If $\G$ is rigid, then $girth(G_n) = \omega(1)$. 
On the other hand,  if  $girth(G_n) = \Theta(\log{n})$  and the minimum check degree of $\G$ is at least $3$ (i.e., for all $n$, each check node of $G_n$ has degree at least $3$), then the following are equivalent: 
\begin{itemize}
\item[i)] {\em (Rigidity)} $\G$ is rigid 
\item[ii)] {\em (Nondegeneracy)} For each constant $c >0$, $G_n$ is $(c \log{n}, \omega(1))$-nondegenerate. \\
That is,  for each constant $c>0$, the minimum weight of a sum of at least $c \log{n}$ checks nodes tends to infinity as $n$ increases.
\end{itemize}
}
\noindent
{\bf Proof:} 
First we show that if $\G$ is rigid, then $girth(G_n) = \omega(1)$. 
If  $G_n$ has  a cycle of $O(1)$ length, then
 the weight of the sum of the check nodes on this cycle is $O(1)$ since 
$\G$ has bounded check degree, which contradicts the rigidity of $\G$.

Assume in what follows that: 
\begin{itemize}
\item[a)] $girth(G_n) = \Theta(\log{n})$ and let $\alpha>0$ be a constant such that $girth(G_n) \geq \alpha \log{n}$ for 
sufficiently large $n$.  
\item[b)] Each check node of $G_n$ has degree at least $3$.  
\end{itemize}
Say that $G_n = (V_n,C_n,E_n)$, let $Q_n$ be the code associated with $G_n$ and 
let $z_j\in Q_n^\bot$  be   the vector in the dual code associated with check $j\in C_n$.    
We will use (a) to show that (ii) implies (i) and (b) to show that (i) implies (ii).

Assume that (ii) holds. To verify (i), let
$z= \sum_{j \in S} z_j$ for some subset $S\subset C_n$ such that  the graph induced by $G_n$ on $S$ contains a cycle. 
Thus  $|S|\geq \frac{1}{2}girth(G) \geq \frac{\alpha}{2}\log{n}$.
Since $G_n$ is $(\frac{\alpha}{2}\log{n}, \omega(1))$-nondegenerate, we get 
$weight(z)= \omega(1)$, hence $\G$ is rigid.  

Finally, assume that $\G$ is rigid and let $c>0$. To verify that (ii) holds, we use (b). 
Let $ z= \sum_{j \in S} z_j$ for some subset $S\subset C_n$ of size at least $c\log{n}$.
We will argue that $weight(z)=\omega(1)$ by considering two cases depending 
on whether or not the graph $G_S$ induced by $G_n$ on $S$ is  acyclic. 

Case 1: Assume that $G_S$ is  acyclic. Since each check node 
in $S$ has degree at least $3$, the number of leaves in the forest $G_S$ is at least $|S|+2$ 
(in general, if $F$ is a forest and $s$ is the number of internal nodes of $F$ of degree at least $3$, 
then the number of leaves of $F$ is at least $s+2$ assuming that $s\geq 1$).
Since each of those leaves must be a variable node, we get $weight(z)\geq |S|+2=\Omega(\log{n})$.
   
Case 2: If $G_S$  contains a cycle, then  $weight(z)=\omega(1)$ since 
$\G$ is rigid. 
\finito 
\begin{definition}[\cite{calkin}] \label{calkinthr}{\em\bf (Calkin's threshold)}
If $d\geq 3$ is an integer,  define the threshold  $0 < \beta_d < 1$ as follows. 
 Consider the function  
\[
f_d(\alpha,\beta) = -1+H(\alpha)+\beta\log_2{(1+(1-2\alpha)^d)}, 
\]
where $H(\alpha) = - \alpha\log_2{\alpha} - (1-\alpha)\log_2(1-\alpha)$ is the binary entropy 
function.  Let \[\mbox{$\beta_d := \sup\{ \beta^*$ : $f_d(\alpha, \beta)<0$ for all $0 < \alpha < 1/2$ and all $0<\beta<\beta^* \}.$}\]
Equivalently, $\beta_d$ is the unique $0 < \beta_d < 1$ such that there exists 
$0 < \alpha_d < 1/2$ such that $(\alpha_d, \beta_d)$ is a root of the system of equations   
\[
    \left\{\begin{array}{rll} f_d(\alpha,\beta) &=& 0 \\ \frac{\partial}{\partial \alpha} f_d(\alpha,\beta) &=& 0.   \end{array}\right. 
\]   
\end{definition}
For instance,  $\beta_3 \sim 0.8895$,  $\beta_4 \sim 0.967$ and $\beta_5 \sim 0.989$. As $d$ increases, $\beta_d$ approaches $1$. 
In general,  
Calkin 
shows that $\beta_d = (1  - \frac{e^{-d}}{\ln{2}})(1\pm o(1))$. Calkin established the following. 

\begin{lemma}[\cite{calkin}]\label{calkinthrthm}{\em\bf (Random row-regular matrices have full row rank)}
Let $d \geq 3$ be an integer. Consider a random $m\times n$ matrix $M\in \F_2^{m\times n}$ constructed   by independently 
choosing each of the  $m$ rows of $M$ uniformly from  the set of vectors in $\F_2^n$ of weight $d$.
If  $m < \beta_d n$, then  
the probability that the rows of $M$ are linearly dependent goes to zero as $n$ tends to infinity.  
\end{lemma}
Note that full row rank corresponds to $(1,0)$-nondegeneracy. ~~\\

\noindent 
{\bf Lemma \ref{randlem}}~{\bf  (Random check-regular graphs are nondegenerate)} 
{\em 
Let $d,m$  and $n$ be integers such that $d\geq 3$ and  $1\leq m < \beta_d n$.
Consider a random $m\times n$ matrix $M\in \F_2^{m\times n}$ constructed by independently 
choosing each of the  $m$ rows of $M$ uniformly from  the set of vectors in $\F_2^n$ of weight $d$. 
Then for any constant $c>0$ and any function $k(n)$ of $n$ such that  $k(n) = o(\log{\log{n}})$, 
$M$ is $(c\log{n}, k(n))$-nondegenerate with high probability. 
That is, the probability that there are at least 
$c\log{n}$ rows of $M$ whose $\F_2$-sum has weight less than or equal to $k(n)$ goes to zero as $n$ tends to infinity. 
}

\subsection{Proof of Lemma \ref{randlem}}\label{appa}

The proof follows the argument Calkin \cite{calkin} used to establish Lemma \ref{calkinthrthm}.
Let $B_d$ be the set of  vectors in $\F_2^n$ of weight $d$.
Let $g = \lceil c\log{n} \rceil$, $k=k(n)$, and  $P$ be the probability that there are at least $g$  rows of $M$ whose $\F_2$-sum has weight less than or equal to $k$. Thus 
\begin{equation}\label{pmessexp}
P \leq \sum_{t = g}^m \binom{m}{t} \sum_{p = 0}^k a_p^{(t)},
\end{equation}
where   $a_p^{(t)}$ is the probability that the weight of the sum of $t$ random vectors chosen uniformly and independently from 
$B_d$ is $p$.

Consider the random walk on $\F_2^n$ which starts from $0$ and moves by adding random elements from $B_d$. The transition 
probability matrix of the underlying Markov chain is the $(n+1)\times (n+1)$ matrix $A = (a_{pq})_{p,q\in \{0,\ldots,n\}}$, where 
 $a_{pq}$ is  defined as follows. Fix any vector $y_q\in \F_2^n$ of weight $q$.  Then $a_{pq}$ is 
the probability that the weight of $x+y_q$ is $p$ over the 
uniformly random choice of  $x$ from $B_d$.  
The entries of $A$ are given by 
 $$a_{pq} = \frac{\binom{q}{\frac{q+d-p}{2}}\binom{n-q}{\frac{d-q+p}{2}}}{\binom{n}{d}}$$ if $q+d-p$ is even. Otherwise, $a_{pq}=0$. 
In terms of $A$, $a_p^{(t)} = a_{p0}^{(t)}$, where 
$a_{p0}^{(t)}$ is the $(p,0)$'th entry of the matrix $A^t$. 

The following lemma due to Calkin gives the eigenvalues and the eigenvectors of $A$ in terms of Krawtchook Polynomials.
\begin{lemma}[\cite{calkin}]  \label{calkinlemco}
\begin{itemize}
\item[a)]
The eigenvalues of $A$ are 
\[
     \lambda_i = \frac{1}{\binom{n}{d}}\sum_{s} (-1)^s \binom{i}{s}\binom{n-i}{d-s} ~~~\mbox{for $i = 0,\ldots,n$.}
\] 
The eigenvector corresponding to $\lambda_i$ is the $n\times 1$ vector $e_i$ whose entries are given by 
\[
     e_{ij} = \sum_{s} (-1)^s \binom{i}{s}\binom{n-i}{j-s} ~~~\mbox{for $j = 0,\ldots,n$.}
\] 
Moreover, $A$  is decomposable as $A = U \Lambda U^{-1}$, where $\Lambda = diag(\lambda_i)_{i=1}^n$, $U$  is the matrix whose columns are 
$e_0,\ldots, e_n$ and  $U^{-1} = 2^{-n}U$. 
\item[b)] If $i>\frac{n}{2}$, then $\lambda_i = (-1)^d \lambda_{n-i}$.
\end{itemize}
\end{lemma} 
We have $$a_{p}^{(t)}=a_{p0}^{(t)} = 2^{-n} \sum_i e_{ip}\lambda_i^t e_{0i} \leq 2^{-n} \binom{n}{p}\sum_i \binom{n}{i}  |\lambda_i|^t,$$ 
since $e_{0i}=\binom{n}{i}$ and $|e_{ip}|\leq \binom{n}{p}$.  
It follows from (\ref{pmessexp}), that 
\begin{eqnarray}
P &\leq& 2^{-n} \sum_{p = 0}^k  \binom{n}{p} \sum_{i=0}^n \binom{n}{i}\sum_{t =g}^{m} \binom{m}{t}   |\lambda_i|^t\nonumber\\
  &\leq& 2(n+1)^k\sum_{i=0}^{\floor{n/2}} 2^{-n} \binom{n}{i}  \sum_{t =g}^{m}  \binom{m}{t} |\lambda_i|^t, \label{Pexpmess}
\end{eqnarray}
where the second inequality follows from   Part (b) of Lemma \ref{calkinlemco} 
and the bound $\sum_{p = 0}^k  \binom{n}{p}\leq (n+1)^k$.  
Instead of (\ref{Pexpmess}),  Calkin obtains the bound:   
\begin{equation}\label{Pexpmessc}
      2\sum_{i=0}^{\floor{n/2}} 2^{-n} \binom{n}{i}  \sum_{t =1}^{m}  \binom{m}{t} \lambda_i^t.
\end{equation}
The key differences between   (\ref{Pexpmess}) and  (\ref{Pexpmessc})  are that 
(\ref{Pexpmessc}) 
starts from $t = 1$ instead of $t = g$ and 
(\ref{Pexpmess}) has    the extra  $(n+1)^k$ term (the fact that the absolute values of the eigenvalues appear in (\ref{Pexpmess}) instead of their actual values is of minor 
significance).   
We will show that $P \leq 2^{-\Theta(n^{1/7})}  + \frac{2(n+1)^km}{g^g}$, hence 
$P = o(1)$ for $g = \Theta(\log{n})$ and $k = o(\log{\log{n}})$.

To estimate  $P$, we will use the following bounds on the eigenvalues established by Calkin.
\begin{lemma}[\cite{calkin}]  \label{calkinlemc}
\begin{itemize}
\item[a)] $|\lambda_i|\leq 1$ for all $0 \leq i\leq n$
\item[b)] If  $c n \leq i \leq \frac{n}{2}$ for some constant $c>0$, then 
\[
\lambda_i = \left(1-\frac{2i}{n}\right)^d - \frac{4\binom{d}{2}}{n}
\left(1-\frac{2i}{n}\right)^{d-2}\frac{i}{n}\left(1 - \frac{i}{n}\right) + O\left(\frac{d^3}{n^2}\right).
\]  
\item[c)] If $\frac{n}{2}-n^{4/7} \leq i \leq \frac{n}{2}$, 
then $|\lambda_i| =  o\left(\frac{1}{n}\right)$.
\end{itemize}
\end{lemma}
Let 
\[
P_i =   2(n+1)^k2^{-n} \binom{n}{i}  \sum_{t =g}^{m}  \binom{m}{t} |\lambda_i|^t.
\]
Thus $P \leq \sum_{i=0}^{\floor{n/2}} P_i$. 
We divide the summation on $i$ as in the argument of  Calkin into three regions:    $0\leq i\leq \e n$, $\e n < i\leq n-n^{4/7}$  and  $\frac{n}{2}-n^{4/7}< i \leq \frac{n}{2} $, 
where $\e>0$ is a sufficiently small constant. We will use the condition  $m < \beta_d n$ in second region   
 and fact that  $t$ starts from $g$ in the third region.

{\em Region 1:   
$0\leq i\leq \e n$.}  Using the bound $|\lambda_i|\leq 1$ and ignoring the lower bound $g$ on $t$, we get 
\begin{eqnarray*}
P_i &\leq &  2(n+1)^k2^{-n} \binom{n}{i}  \sum_{t =0}^{m}  \binom{m}{t} |\lambda_i|^t = 2(n+1)^k2^{-n} \binom{n}{i} (1+|\lambda_i|)^m \\
&\leq& 2(n+1)^k\binom{n}{i} 2^{-(n-m)} \leq 2(n+1)^k2^{-n(1-m/n-H((\e)) + O(\log{n})}=2^{-\Theta(n)}, 
\end{eqnarray*}
for sufficiently small $\e>0$,  since $m < \beta_d n$, $\beta_d < 1$ and $k=o(\frac{n}{\log{n}})$.  Hence 
\[
    P^{(1)} :=\sum_{0\leq i\leq \e n} P_i \leq 2^{-\Theta(n)}.
\]

{\em Region 3:  
$\frac{n}{2}-n^{4/7}< i \leq \frac{n}{2} $.} 
Here we use the bound $\lambda_i = o\left(\frac{1}{n}\right)$ in Part (c) of Lemma \ref{calkinlemc}
and the bound  $\binom{m}{t}\leq \left(\frac{em}{t}\right)^t$: 
\begin{eqnarray*}
  P_i &\leq &   2(n+1)^k2^{-n} \binom{n}{i}  \sum_{t =g}^{m}  \left(\frac{em}{t}|\lambda_i|\right)^t\\
 &\leq&  2(n+1)^k2^{-n} \binom{n}{i}  \sum_{t =g}^{m} \frac{1}{t^t}\\   
&\leq& 2(n+1)^k2^{-n} \binom{n}{i} \frac{m}{g^g},
\end{eqnarray*}
where the second equality holds for sufficiently large $n$. 
Hence 
 \[
  P^{(3)} := \sum_{\frac{n}{2}-n^{4/7}< i \leq \frac{n}{2}} P_i \leq \frac{2(n+1)^km}{g^g}.
\]

{\em Region 2:  $\e n < i\leq \frac{n}{2}-n^{4/7}$.}  
As in the first region, 
\[
P_i \leq   2(n+1)^k2^{-n} \binom{n}{i}  \sum_{t =0}^{m}  \binom{m}{t} |\lambda_i|^t = 2(n+1)^k2^{-n} \binom{n}{i} (1+|\lambda_i|)^m 
\]
Now, we use 
the bound on $\lambda_i$ in Part (b) of Lemma \ref{calkinlemc} which implies that 
\[
  |\gamma_i| \leq  \left(1-\frac{2i}{n}\right)^d +O\left(\frac{d^3}{n^2}\right). 
\]
Thus 
\[
   (1+|\lambda_i|)^m \leq \left(1+\left(1-\frac{2i}{n}\right)^d +O\left(\frac{d^3}{n^2}\right)\right)^m 
=\left(1+\left(1-\frac{2i}{n}\right)^d\right)^m (1+ o(1)).
\]
For the binomial coefficients, we use the bound  $\binom{n}{i} \leq \frac{e}{2\pi\sqrt{\e(1-\e)n}}2^{n H(\frac{i}{n})}$ which holds for  $\e n \leq i\leq n-\e n$ and follows from Stirling's approximation.
It follows that 
\[    P_i \leq \delta n^{k-\frac{1}{2}}2^{-n(1-H(\frac{i}{n}))}\left(1+\left(1-\frac{2i}{n}\right)^d\right)^m,
\]
for some absolute constant $\delta>0$ and sufficiently large $n$.  Therefore 
 \begin{eqnarray*}
  P^{(2)}&:=& \sum_{\e n < i \leq \frac{n}{2}-n^{4/7}} P_i\\
 &\leq& \delta  n^{k-\frac{1}{2}}\sum_{\e n < i \leq \frac{n}{2}-n^{4/7}} 2^{-n(1-H(\frac{i}{n}))}\left(1+\left(1-\frac{2i}{n}\right)^d\right)^m\\
&=& \delta  n^{k-\frac{1}{2}}\sum_{\e n < i \leq \frac{n}{2}-n^{4/7}} 2^{nf_d(\frac{i}{n},\frac{m}{n}) }.  
 \end{eqnarray*}
By the definition of $\beta_d$, we have $f_d(\frac{i}{n},\frac{m}{n})<0$ for all $\e n < i \leq \frac{n}{2}-n^{4/7}$  since $\frac{m}{n}< \beta_d$. 
Moreover, since $f_d(\frac{1}{2},\beta)=0$ for each $\beta$, the maximum of $f_d(\frac{i}{n},\frac{m}{n}) $ over $\e n < i \leq \frac{n}{2}-n^{4/7}$  occurs at 
$i = \lfloor \frac{n}{2}-n^{4/7} \rfloor$. It follows that 
\[
  P^{(2)} \leq \delta  n^{k+\frac{1}{2}} 2^{nf_d(\frac{\lfloor \frac{n}{2}-n^{4/7} \rfloor}{n},\frac{m}{n})  }.       
\]
For $|\alpha-\frac{1}{2}|= o(1)$ and $\beta>0$, 
$f_d(\alpha,\beta) = -1+H(\alpha)+\beta\log_2{(1+(1-2\alpha)^d)} = - \Theta((\alpha - \frac{1}{2})^2)$
since  $H(\alpha) = 1 - \Theta((\alpha-\frac{1}{2})^2)$ and $d\geq 3$. It follows that $P^{(2)} \leq \delta  n^{k+\frac{1}{2}} 2^{-\Theta(n^{1/7})}  = 2^{-\Theta(n^{1/7})}$  
for $k  =o\left(\frac{n^{1/7}}{\log{n}}\right)$.

Combining the above three cases, we get 
\[
P \leq P^{(1)}+P^{(2)}+P^{(3)}\leq 2^{-\Theta(n^{1/7})}  + \frac{2(n+1)^km}{g^g}
\]
if  $k  =o\left(\frac{n^{1/7}}{\log{n}}\right)$.  Recall that $g = \Theta(\log{n})$. 
It follows that $P = o(1)$ if $k = o(\log{\log{n}})$.

\section{Pseudocodewords  interpretation of asymptotic strength}\label{pseudoint}
   
In the section we give an interpretation of the notion of asymptotic strength  in terms of the 
fractional  spectrum of pseudocodewords. Then we compare with the related notions of minimum BSC-pseudoweight \cite{KVLarge},   
 fractional distance  and maximum-fractional distance \cite{Fel03,FWK05}.

If $G = (V,C,E)$ is a Tanner graph, let $Ext(G)$ be  the set of extreme points  of $P(G)$.  
The codewords of $Q$ are the integral vertices of $P(G)$, 
i.e., $Ext(G)\cap \{0,1\}^n = Q$.  The elements of $Ext(G)$ are called  {\em pseudocodewords} (see \cite{Fel03,KV03,FWK05,KVLarge}).

 In terms of pseudocodewords,  the notion  of asymptotic strength 
translates as follows.

\begin{theorem}[Pseudocodewords and asymptotic strength]\label{asseq}
Let $\G = \{ G_n\}_n$ be an infinite family of Tanner graphs. 
Then $\G$ is asymptotically strong iff for each  (small) constant $\theta>0$, there exists a constant 
$\alpha>0$ such that for each $n$ and each nonzero pseudocodeword $x\in Ext(G_n)$, 
the sum of the largest $\lfloor \alpha n \rfloor $ coordinates  of $x$ is less than $\theta\sum_i x_i$. 
That is, to attain a  positive fraction of $\sum_i x_i$, we need a least  linear number of coordinates of $x$.  
\end{theorem}

\noindent 
{\bf Proof:} 
By the definition of the LP decoder, the following are equivalent for any LLR vector $\gamma \in \R^n$:   
\begin{enumerate}
\item[i)] The LP decoder of $G_n = (V_n,C_m,E_n)$ succeeds on $\gamma$ under the all-zeros codeword assumption 
\item[ii)] $\langle x, \gamma\rangle >0$  for each nonzero pseudocodeword  $x\in Ext(G)$. 
\end{enumerate}
By the equivalence between (i) and (ii), 
 $\G$ is asymptotically strong iff for each   constant $\beta>0$, there exists a constant 
$\alpha>0$ such that for each $n$ and each error vector  $y\in \{0,1\}^n$ of weight at most $\alpha n$, we have 
$\langle x, \gamma({y,\beta}) \rangle >0$  for each nonzero pseudocodeword  $x\in Ext(G_n)$, where  
$\gamma(y,\beta): V_n \rightarrow \R$ is the asymmetric LLR vector  given by 
\[                        
\gamma_i(y,\beta) = \left\{\begin{array}{ll} -1 &\mbox{ if } y_i=1  \\ \beta & \mbox{ if } y_i=0.\end{array}\right.
\]
Let $U = \{ i: y_i =1\}$, thus 
\[
\langle x, \gamma({y,\beta}) \rangle   = - \sum_{i\in U } x_i +  \beta\sum_{i\in U^c} x_i 
=- (1+\beta)\sum_{i\in U } x_i +  \beta \sum_{i} x_i.
\]
Hence $\langle x, \gamma({y,\beta})\rangle >0$  is equivalent to   $\sum_{i\in U } x_i < \frac{\beta}{1+\beta}\sum_{i} x_i$. 
The  theorem then follows by setting $\theta = \frac{\beta}{1+\beta}$.
\finito

Note that if $x$ is integral, i.e.,  $x\in \{0,1\}^n$ is a codeword, then the above condition is equivalent to 
$weight(x) = \Theta(n)$.  If $x$ is not integral,  the above condition says that the fractional weights 
spectrum is not ``too unbalanced''  in the sense that we need at least a linear number of coordinates of $x$ to attain a  positive fraction of $\sum_i x_i$.

In the setup of Theorem \ref{asseq}, 
the minimum BSC-pseudoweight \cite{KVLarge}  $w^{BSC}_p(G_n)$ corresponds to $\theta = \frac{1}{2}$. 
Namely, $w^{BSC}_p(G_n) = 2 a^*$, where $a^*$ is the maximum value of $a$ such that the 
sum of the largest $a$ 
coordinates  of $x$ is less than $\frac{1}{2}\sum_i x_i$ for all nonzero $x\in Ext(G_n)$. 
The $2$ multiplicative factor ensures that the largest number of errors the LP decoder can handle over the BSC is  $w^{BSC}_p(G_n)/2$.
Thus, for integral codewords,  the BSC-pseudoweight coincides with  the Hamming weight.  
The asymptotic strength property implies that  $w^{BSC}_p(G_n) = \Theta(n)$. 
It is not clear if the converse holds; 
the  asymptotic strength requirement seems stronger since it is in terms of all  $\theta> 0$ and not only $\theta = \frac{1}{2}$.
We leave the problem of whether or not it  is strictly stronger open. 

The fractional distance of $G$ is the minimum $L_1$-norm of a nonzero  pseudocodeword \cite{Fel03,FWK05}. 
Unlike the 
 the minimum BSC-pseudoweight,  the   fractional distance is always  sublinear for regular bounded-degree 
Tanner graphs 
 \cite{KV03,KVLarge}. 
 The same holds for the maximum-fractional distance 
which  is defined as the minimum $L_1$-norm/$L_{\infty}$-norm of a nonzero  pseudocodeword \cite{Fel03,FWK05}.

\section{Decoding with help bits}\label{dechelp}

In this section we highlight a general property 
of asymptotically strong Tanner graphs. We argue 
that for such graphs, allowing 
a sublinear number of ``help bits'' does not improve the LP threshold. 
This result,  although a negative statement, has potential constructive applications as it weakens the dual witness 
requirement for LP decoding success. 
We also derive a converse of the LP excess lemma.

\begin{definition}[LP decoder with help] 
Let $\H=\{ H_n\}_n$ be an  infinite   family of  Tanner graphs and $b:\N \rightarrow \R^{\geq 0}$.
Consider transmitting  $x\in \F_2^n$ and receiving the corrupted version  $y\in \F_2^n$ of $x$. 
We say that the LP decoder of $H_n$     {\em corrects $y$ with $b(n)$ help bits} if 
there exists $z\in \F_2^n$ of weight at most $b(n)$ such that the LP decoder  of $H_n$ 
succeeds in recovering $x$ from  $y+z$. That is, we are allowed to flip at most $b(n)$ bits of $y$ to help 
the LP decoder. 
Define the {\em LP-threshold $\xi_{LP}(\H,b)$}   to be 
the supremum of $\e\geq 0$ such that the probability that the LP decoder of $H_n$ 
 fails with $b(n)$ help bits   over the $\e$-BSC tends to zero as $n$ tends to infinity, i.e.,  
\[
   \xi_{LP}(\H,b) = \sup\{\e\geq 0   ~:~  \operatorname{Pr}_{\e\mbox{-}BSC}[\mbox{LP decoder  of $H_n$ fails with $b(n)$ help bits}] = o(1) \}.  
\]
\end{definition}

\begin{theorem}[Sublinear help does not improve LP threshold]  \label{tbcp2}
Let $\H=\{H_n\}_n$ be an  infinite   family of  Tanner graphs. 
If $\H$ is asymptotically strong and $b(n) = o(n)$, then $\xi_{LP}(\H,b)  = \xi_{LP}(\H)$.
\end{theorem} 
A potential constructive application of Theorem \ref{tbcp2} is the following. 
In dual terms (by Theorem \ref{crr1}), the LP decoder of $H_n = (V_n,C_n,E_n)$     corrects $y$ with $b(n)$ help bits  iff there is  a $b(n)$-weak dual witness 
for $(-1)^y$, where $w: V\rightarrow \R$ is called a {\em $b(n)$-weak dual witness}  if  
instead of the variable nodes inequalities $F_i(w) < (-1)^{y_i}$, for $i\in V$,   
it satisfies  the  following  weaker version: 
 \begin{equation*}
\left\{\begin{array}{lllll} F_i(w) &<& 1 &\mbox{ for all } i\in V_n  \\ 
 F_i(w) &<& -1&\mbox{ for all but at most $b(n)$ variable $i\in V_n$ such that $y_i = 1$.}  
\end{array}\right.
\end{equation*}
Thus Theorem \ref{tbcp2} implies that  to estimate the LP threshold of an asymptotically strong Tanner graph, it is 
enough to find a weak dual witness instead of a dual witness, 
which is in principle an easier task.

The proof of Theorem \ref{tbcp2} is below and it uses the following converse of the LP excess lemma (Lemma \ref{excesslem}).
\begin{lemma}[LP deficiency lemma: trading LP deficiency with crossover probability]  \label{defilemma} 
Let $H = (V,C,E)$ be a Tanner graph.  
Let $0<\e<\e'<1$ and $0< \delta< 1$ such that $\epsilon' = \epsilon + (1-\epsilon)\delta$. 
Let $q_{\epsilon',\delta}$ be the probability that there is no dual witness in $H$ for $(-1)^y+\frac{\delta}{2}$, where $y \sim \operatorname{Ber}(\epsilon',n)$ 
is an  error pattern generated by the $\e'$-BSC. 
Then the probability that the   LP decoder of $H$ fails on the $\e$-BSC is at most $\frac{2q_{\epsilon',\delta}}{\delta}$.  
\end{lemma}
The proof of the LP deficiency lemma
is in Section \ref{defilemmap}. Note that the $\frac{\delta}{2}$ term in $(-1)^y+\frac{\delta}{2}$ represents the ``LP deficiency'' of the dual witness with respect to $(-1)^y$, i.e., how far it is from being a dual witness for $(-1)^y$.

~\\
\noindent 
{\bf Proof of Theorem \ref{tbcp2}:}
The proof uses a part of the  argument in Theorem \ref{crr1} and applies  the LP deficiency lemma instead of the LP excess lemma. 
Using the asymptotic strength of $\H$, we will  trade the help bits with LP deficiency, which in turns we will trade with 
crossover probability using   the LP  deficiency lemma.
At a high level, the argument is as follows. 
For any $\delta> 0$, 
we will  operate the LP decoder of 
${\H}$ with $b(n)$ help bits on the BSC below its threshold $\xi_{LP}({\H},b)$  by around $\frac{\delta}{2}$. 
With high probability, we have a  dual witness $w$ for $(-1)^{y+z}$ for some help vector 
$z\in \{0,1\}^n$ of sublinear weight. We will turn $w$ into a dual witness for $(-1)^y + \frac{\delta}{4}$ by 
patching  to $w$ a dual witness $v$ for the asymmetric LLR vector $\mu({z,\delta})$ given by 
 \begin{equation}\label{sddsss}                        
\mu_i({z,\delta}) = \left\{\begin{array}{ll} -2 &\mbox{ if } z_i=1  \\ \frac{\delta}{4} & \mbox{ if } z_i=0,\end{array}\right.
\end{equation}
for all $i\in V$. The existence of $v$ follows from the asymptotic strength of $\H$.
Using the LP deficiency lemma, we get rid of the deficiency  $\frac{\delta}{4}$ by deceasing the crossover probability 
to $\xi_{LP}({\H},b)-\delta$.

More precisely, assume without loss of generality that $\xi_{LP}({\H},b)>0$ and 
consider any (small) constant $0< \delta < \xi_{LP}({\H},b)$. 
We will show that       $\xi_{LP}({\H})\geq \xi_{LP}({\H},b)-\delta$. 
Let  $\e = \xi_{LP}({\H},b)-\delta$ and $\e' = \e+(1-\e)\frac{\delta}{2}$, thus $0<\e<\e'< \xi_{LP}({\H})$. 
 Let $q_{\e'}(n)$ be the probability that the LP decoder of  ${H_n}$ with $b(n)$ help bits fails on $(-1)^y$, where 
 $y \sim \operatorname{Ber}(n,\e')$. 
Since  $\e'< \xi_{LP}({\H},b)$, we have $q_{\e'}(n)  = o(n)$.   
By Theorem \ref{equivChars}, with probability $1-q_{\e'}(n)$ 
over the choice of  $y \sim \operatorname{Ber}(n,\e')$, 
there  is  a dual witness $w$ in ${H_n}$ for $(-1)^{y+z}$ for some 
$z\in \{0,1\}^n$ of weight at most  $b(n)$. 
Consider any $n$ and  any $y\in \{0,1\}^n$ such that $w$ and $z$ exist. 
Since $\H$ is asymptotically strong, there exists a constant 
$\alpha_{\delta} >0$ (independent of $n$) such that if $weight(z)\leq \alpha_\delta n$, 
 the LP decoder of $H_n = (V,C,E)$ succeeds on the 
asymmetric LLR vector $\mu({z,\delta})$ defined in (\ref{sddsss}).  
Accordingly, by Theorem \ref{equivChars}, let $v: E \rightarrow \R$ be a dual witness for $\mu({z,\delta})$.
Since $b(n) = o(n)$, assume that $n$ is large enough so that $b(n) \leq \alpha_{\delta} n$.
It follows that $w+v$ is a dual witness for $(-1)^{y+z}+\mu({z,\delta})$. Since 
$(-1)^{y+z}+\mu({z,\delta}) \leq (-1)^y + \frac{\delta}{4}$, we get 
that  $w+v$ is a dual witness for  $(-1)^y + \frac{\delta}{4}$. 
Therefore,  the probability that there is no dual witness  in 
${H_n}$ for $(-1)^y + \frac{\delta}{4}$  over the choice 
of $y \sim \operatorname{Ber}(n,\e')$ is at most  $q_{\e'}(n)$.  
It follows from the LP deficiency lemma that  the probability that the   LP decoder of 
$H_n$ fails on the $\e$-BSC is at most $\frac{4q_{\epsilon'}(n)}{\delta}$.  
Since $q_{\e'}(n)  = o(n)$, we get that $\xi_{LP}({\H})>\e = \xi_{LP}({\H},b)-\delta$.
\finito

\subsection{Proof of Lemma \ref{defilemma}} \label{defilemmap} 
The proof is a variation of the  argument in  the proof of Theorem 8.1 in \cite{BGU14}.  
Decompose the $\epsilon'$-BSC into the bitwise OR of the $\epsilon$-BSC and the $\delta$-BSC. Choose 
$x \sim \operatorname{Ber}(\epsilon,n)$ and  $e'' \sim \operatorname{Ber}(\delta,n)$  and consider $e = x \lor e''$, thus  $e \sim \operatorname{Ber}(\epsilon',n)$. 
At a high level, we will construct a dual witness on the $\e$-BSC by appropriately averaging 
dual witnesses on the $\e'$-BSC over the choice of $e'' \sim \operatorname{Ber}(\delta,n)$.

For every $y \in \{0,1\}^{n}$, let 
\[
L(y) = \left\{\begin{array}{ll}1 & \mbox{if  $(-1)^y+\frac{\delta}{2}$ has a dual witness}\\ 0 &\mbox{otherwise.}\end{array}\right.
\]
Thus, in terms of $L$, 
\begin{equation}\label{pyly}
q_{\epsilon',\delta} = Pr_{y \sim \operatorname{Ber}(\epsilon',n)}\big[L(y)=0].  
\end{equation}
If $y \in \{0,1\}^n$, let  $v^{y}: E\rightarrow \R$ be an arbitrary dual witness for $(-1)^y+\frac{\delta}{2}$ if $L(y)=1$. Otherwise, let $v^y:E \rightarrow \R$ 
be the identically zero function. If $x \in \{0,1\}^n$, define 
$w^x: E \rightarrow \R$ by averaging $v^{x \lor e''}$ over the choice of $e'' \sim \operatorname{Ber}(\delta,n)$ and scaling: 
 $$w^x =\alpha \operatorname{E}_{e'' \sim \operatorname{Ber}(\delta,n)} v^{x \lor e''},$$ 
where $\alpha = \frac{1}{1-\delta}>0$. 
We will show that $w^x$ is a dual witness for $(-1)^x$ with probability at least $1-\frac{2q_{\e',\delta}}{\delta}$ over the choice of $x\sim \operatorname{Ber}(\e,n)$.

If $L(y)=1$,   then by definition,  $v^y$ satisfies the dual witness check nodes inequalities: 
$v^y(i,j) + v^y(i',j) \geq 0$, 
for each check $j\in C$  and all distinct variables $i\neq i' \in N(j)$. 
The identically zero function $E \rightarrow \R$ also satisfies those  inequalities, hence they are satisfied by 
$v^y$ for all $y\in \{0,1\}^n$. Since $w^x$ is an average over  $v^{x\lor e''}$ scaled by a positive constant, we get that 
the dual witness check nodes inequalities  are satisfied by $w^x$ for all $x\in \{0,1\}^n$.

In what follows, we  take care of the variable nodes inequalities  ((a) in Definition \ref{dualdefs}).  
If $w: E\rightarrow \R$, consider the flow vector  $F(w)\in \R^V$ associated with $w$:  
$
            F_i(w) = {\displaystyle\sum\limits_{j \in N(i)} w(i,j)},  
$
for all $i\in V$.  In terms of $F$, we have  
\begin{equation}\label{fvymy}
F(v^y)< (-1)^y+\frac{\delta}{2}  \text{, for each }y\in \{0,1\}^n\mbox{ such that }L(y)=1.
\end{equation}
We have to show that $F(w^x) < (-1)^{x}$ with probability at least $1-\frac{2q_{\e',\delta}}{\delta}$ over the choice of $x\sim \operatorname{Ber}(\e,n)$.  
For any $x\in \{0,1\}^n$, 
\begin{eqnarray}
F(w^{x})&=&\alpha \operatorname{E}_{e''\sim \operatorname{Ber}(\delta,n)} F(v^{x \lor e''})\nonumber \\ 
&=& \alpha \operatorname{E}_{e''}[~F(v^{x \lor e''})~|~L(x \lor e'')=1~]\times \operatorname{Pr}_{e''} [L({x \lor e''})=1]\label{eqavw}\\ 
&<& \alpha \operatorname{E}_{e''}[~(-1)^{x \lor e''}+\mbox{$\frac{\delta}{2}$}~|~L({x \lor e''})=1~]\times \operatorname{Pr}_{e''}[L({x \lor e''})=1]~~~\text{(using (\ref{fvymy}))}\nonumber \\ 
&=&\alpha \Big(\operatorname{E}_{e''}(-1)^{x \lor e''} +\mbox{$\frac{\delta}{2}$}- E_{e''}[~(-1)^{x \lor e''}~|~L({x \lor e''})=0~] \times \phi_{x}\Big)\nonumber \\ 
&\leq& \alpha \Big(\operatorname{E}_{e''}(-1)^{x \lor e''} +\mbox{$\frac{\delta}{2}$}+ \phi_{x}\Big)\nonumber
\end{eqnarray}
where $\phi_{x} := \operatorname{Pr}_{e''\sim \operatorname{Ber}(\delta,n)}\big[L({x \lor e''})=0]$. 
Note that (\ref{eqavw}) follows from the fact that $L(y)=0$ implies $v^y=0$ and hence $F(v^y)=0$.
Fix any  $i \in V$. 
If  $x_{i}=1$, then $\operatorname{E}_{e''} (-1)^{x_i \lor e''_i} = -1$. 
 If $x_{i}=0$,  then 
$\operatorname{E}_{e''}(-1)^{x_i \lor e''_i} = \delta(-1)+(1-\delta)(1) = 1-2\delta$. Hence 
\[
 F_{i}(w^{x}) < \begin{dcases*}
        \alpha(-1+\mbox{$\frac{\delta}{2}$}+\phi_{x})  & if $x_{i}=1$\\ 
        \alpha (1-\mbox{$\frac{3\delta}{2}$}+\phi_{x}) & if $x_{i}=0$.
        \end{dcases*}
\]
By (\ref{pyly}),   
$$\operatorname{E}_{x \sim \operatorname{Ber}(\epsilon,n)} \phi_{x} = \operatorname{Pr}_{e''\sim \operatorname{Ber}(\delta,n),x \sim \operatorname{Ber}(\epsilon,n)}\big[L({x \lor e''})=0]  = 
\operatorname{Pr}_{y\sim \operatorname{Ber}(\epsilon',n)}\big[L(y)=0] = q_{\epsilon',\delta}.$$ Thus, by Markov's inequality,
$\phi_{x} \ge \frac{\delta}{2}$ with probability at most $ \frac{2q_{\epsilon',\delta}}{\delta}$ over the choice $x \sim \operatorname{Ber}(\epsilon,n)$.  Hence, 
with probability at least $ 1-\frac{2q_{\epsilon',\delta}}{\delta}$ 
over $x \sim \operatorname{Ber}(\epsilon,n)$, we have for all $i \in V$,  
\begin{eqnarray*}
 F_{i}(w^{x}) &<& \begin{dcases*}
        \alpha(-1+\mbox{$\frac{\delta}{2}$}+\mbox{$\frac{\delta}{2}$})  & if $x_{i}=1$\\ 
        \alpha (1-\mbox{$\frac{3\delta}{2}$}+\mbox{$\frac{\delta}{2}$}) & if $x_{i}=0$.
        \end{dcases*}\\
          &=& (-1)^{x_i},
\end{eqnarray*}
since $\alpha = \frac{1}{1-\delta}$.

\section{Discussion and open problems} \label{disop}

We conclude with some remarks and open questions mainly related to the asymptotic strength condition, the  rigidity condition and the LP decoding threshold on the BSC.

~\\
{\bf  Asymptotic strength condition.} Theorem \ref{expstr} shows that expansion implies asymptotic strength. 
We know that random low density Tanner graphs are good expanders with high probability \cite{SS96,FMS07}. 
Combining Theorem \ref{expstr} and the probabilistic analysis in Appendix B of \cite{FMS07} implies the following. 
\begin{theorem}\label{exphft}
Let $0< r< 1$ be a constant. Let $d_v$ be a positive  integer  constant such that  
there exists a constant $\frac{2}{3} < \delta < 1$ for which 
$\delta d_v$ and  $(1-\delta)d_v$ are   integers and  $(1-\delta)d_v\geq 2$.   
Then, for any positive integers $n$ and  $m$ such that $r = 1 - \frac{m}{n}$, a random 
variable-regular Tanner graph $G$ with variable degree $d_v$, $n$ variable nodes  and $m$ check nodes  
is asymptotically strong with high probability\footnote{
By a random family $\G = \{G_n\}_n$ of Tanner graphs  being asymptotically strong with high probability, we mean the following. For each constant 
$\beta>0$, there exists a constant  $\alpha>0$ such that for each $n$, with probability at least $1-o(1)$ over the random choice of $G_n$, 
  the LP decoder of $G_n$ succeeds on the asymmetric LLR vector $\gamma({y,\beta})$ for all $y\in \{0,1\}^n$ of weight at most $\alpha n$.}.    
\end{theorem}
The integrality constraint on $\delta d_v$ and $(1-\delta)d_v$ can require large values of $d_v$ (see \cite{FMS07}). 
We conjecture that the following holds.
\begin{conjecture} For all $d_c> d_v\geq 3$, 
a random $(d_v,d_c)$-regular Tanner graph is asymptotically strong with high probability.
\end{conjecture} 

~\\
{\bf Rigidity condition.}   If the graph has $\Theta(\log{n})$ girth and  minimum check degree at least $3$, the rigidity condition 
is equivalent to the simpler $(c\log{n}, \omega(1))$-nondegeneracy condition. We argued in Lemma \ref{randlem}
that the latter condition holds with high probability for   
random check-regular graphs assuming that 
$m< \beta_d n$, where $m$ is the number of  checks nodes, $d$ is the check degree and $\beta_d$ is Calkin's threshold. 
 The statistical independence of the check nodes in the   ensemble of  random check-regular graphs 
makes the ensemble  attractive from a probabilistic analysis perspective, but it  typically gives  irregular graphs with constant girth.
 We believe that good girth and variable-regularity 
  do not increase the odds of   degeneracy; 
  we conjecture that  nondegeneracy is also a typical property of the 
 ensemble of regular  $\Theta(\log{n})$-girth  Tanner graphs.  
\begin{conjecture} 
Let $d_c > d_v \geq 3$ be integers  such that $d_v < \beta_{d_c} d_c$.
 If $\lambda >0$ be a constant,  let $\Gamma_{\lambda }$ be ensemble of $(d_v,d_c)$-regular  Tanner graphs  on $n$ variable nodes 
of girth at least $\lambda \log{n}$.   
Then there is a constant $\lambda >0$ small enough such that  for each constant $c>0$, 
a random graph $G$ from the ensemble $\Gamma_{\lambda }$ is $(c\log{n}, \omega(1))$-nondegenerate with high probability. 
\end{conjecture}
Establishing this conjecture requires working in a more complex probabilistic framework. 
We leave the question open for further investigation.
Note that since $d_c m = d_v n$, the condition $d_v < \beta_{d_c} d_c$ is equivalent to $m < \beta_{d_c} n$. 
A natural but probably more difficult problem is to study also the asymptotic strength of the ensemble $\Gamma_\lambda$.

~\\
{\bf  Limits of LP decoding on the BSC.}
On the positive side,  
our negative results  suggest studying the  LP decoding limits  
in the framework of the dual code containing all redundant checks. 
This framework is appealing since it is independent of a particular Tanner graph representation of the  code.
If $r'$ is the {\em rate of the dual} code, Shannon's limit says that we can transmit reliably over the $\e$-BSC if $\e< H^{-1}(r')$, 
where $H$ is the binary entropy function. For LP decoding with all redundant checks included, it is natural to study 
the following LP capacity function. 
\begin{definition} [LP capacity over the BSC] 
Given a dual rate $0< r'< 1$, define the LP capacity function $$\xi_{LP}(r') : =\underset{\{ D_n\}_n} {\operatorname{\sup}}\xi_{LP}(\{ G_{D_n}\}_n),$$ where the supremum
 is over all $\F_2$-linear codes $D_n\subset \F_2^n$ such that of $\lim_{n\rightarrow \infty} rate(D_n) = r'$ and  
 $G_{D_n}$ is the Tanner graph on $n$ variables whose checks are the nonzero elements of   $D_n$.
\end{definition}
Note that primitive hyperflows (Theorem \ref{primlem})  maybe useful  in studying the LP capacity function.
\begin{question}
\begin{itemize}
\item[i)] {\bf (Relation to Shannon's capacity)} How far is $\xi_{LP}(r')$  from the Shannon's capacity $H^{-1}(r')$? Is $\xi_{LP}(r')=H^{-1}(r')$?
\item[ii)] {\bf (Achievability with bounded check-degree)} Is any   $\e< \xi_{LP}(r')$ achievable by  a family of codes $\{ D_n\}_n$ with a bounded-weight basis? 
That is, is it true that for each  $\e< \xi_{LP}(r')$, there exist a constant $d$ and 
  a family of Tanner graphs $\G = \{G_n\}_n$ such that $\xi_{LP}(\overline{\G}) \geq \e$ 
and $G_n$ has at most $r' n$   check nodes each of degree most $d$?   
\item[iii)] {\bf (Achievability with asymptotic strength)} If the answer of (ii) is affirmative, is $\G$ asymptotically strong? 
\item[iv)] {\bf (Achievability with rigidity)} If the answer of (iii) is affirmative, is $\G$ rigid? 
\end{itemize} 
\end{question}
The answer to first question is not clear. 

The answer to (ii) is probably affirmative since we already know from \cite{FMS07} that a positive value of $\e$ is 
achievable with bounded check degree.    
The answer to (iii) seems also affirmative.
In general,   asymptotic strength makes the LP stronger as it guarantees that the fractional weight spectrum of the pseudocodewords is not ``too unbalanced'' (Theorem \ref{asseq}).  Inspired by \cite{LMSS01}, if $\G$ is not asymptotically strong, we can actually make it 
asymptotically strong without noticeably affecting  its  rate  
 by adding to the code a small  number 
 of parity checks which form 
a sufficiently good  expander \footnote{We need a  $(\varepsilon n, \delta d')$-expander between $n$ variable nodes and $\alpha n$ check nodes 
of  regular variable degree $d'$  and bounded check degree, where $\alpha>0$ is a small constant and $\varepsilon,\delta>0$ are constants such that $\frac{2}{3}< \delta< 1$ and $\delta d'$ is an integer.}.  
The added checks do not decrease the LP threshold of the code.

If both (ii) and (iii) have affirmative answers, we obtain from Theorem \ref{crr1} that
  for any  $\e< \xi_{LP}(r')$, there exists a sufficiently large constant $k\geq d$ such that 
$\xi_{LP}(\overline{\G}^k) \geq \e$. Thus, by running the LP decoder of $\overline{\G}^k$, 
dual rate $r'$ is achievable on the $\e$-BSC  
in time polynomial in the block length $n$. More specifically, the time is polynomial in 
$n^k$, where the constant $k$ increases as the gap $\delta= \xi_{LP}(r')-\e$ gets small. 
 
The last question is more intriguing. 
If the answer to (iv) is also affirmative, then    
$\xi_{LP}(\overline{\G})  = \xi_{LP}(\G)$ by Corollary \ref{crr2.2}. Thus, by running the LP decoder of $\G$, we conclude that 
for any $\e< \xi_{LP}(r')$, dual rate $r'$ is achievable on the $\e$-BSC  
in time polynomial in the block length $n$ and independent of the gap $\delta$, which  is counter intuitive if 
$\xi_{LP}(r')=H^{-1}(r')$.

~\\
{\bf AWGN.} 
On a  final note, a natural question is  to explore the potential extendability of  the results in this paper 
to other channels such as the AWGN.

\section*{Acknowledgments} 
The authors would like to thank the anonymous reviewers for  their helpful and constructive comments.




\end{document}